\documentclass[11pt]{article}

\usepackage{epsfig,psfrag}
\usepackage{color}
\usepackage{array}

\usepackage{amsmath,amsfonts,amssymb}

\renewcommand{\vec}[1]{\boldsymbol{#1}}

\newcommand{\brho}{\bar{\rho}}

\newcommand{\del}{\partial}

\title{A comparison of macroscopic models describing the collective response of sedimenting rod-like particles
in shear flows.}
\author{
Christiane Helzel\thanks{Institute of Mathematics,
  Heinrich-Heine-University D\"usseldorf, D\"usseldorf, Germany. {\tt
    christiane.helzel@hhu.de}} 
\and 
Athanasios E. Tzavaras
\thanks{Computer, Electrical, Mathematical Sciences \& Engineering Division, King Abdullah University of Science and Technology (KAUST), Thuwal, Saudi Arabia.
{\tt athanasios.tzavaras@kaust.edu.sa} }}
\providecommand{\keywords}[1]{\textbf{\textbf{Key words:}} #1}

\date{}

\begin{document}
\maketitle

\begin{abstract}
We consider a kinetic model, which describes the sedimentation of
rod-like particles in dilute suspensions under the influence of
gravity. This model has recently been derived by Helzel and Tzavaras
in \cite{HT2015}. 
Here we restrict our considerations to shear flow and consider a
simplified situation, where the particle orientation is restricted to
the plane spanned by the direction of shear and the direction of gravity.
For this simplified kinetic model we carry out a linear stability
analysis and we derive two different macroscopic
models which describe the formation of clusters of higher particle
density.  One of these macroscopic models is based on a diffusive scaling, the
other one is based on a so-called quasi-dynamic approximation. 
Numerical computations, which compare the predictions of the
macroscopic models with the kinetic model, complete our presentation. 
\end{abstract}


\section{Introduction}
We discuss different mathematical models which describe the
sedimentation process for dilute suspensions of rod-like particles
under the influence of gravity.
The sedimentation of rod-like particles 
has been studied by several authors in theoretical
as well as experimental and numerical works, see Guazzelli and
Hinch \cite{GH11} for a recent review paper.
Experimental studies of Guazzelli and coworkers
\cite{HG96,HG99,MBG07} 
start with a well stirred suspension. Under the influence of gravity, 
a well stirred initial configuration is unstable and it is observed that clusters with
higher particles concentration form. These clusters have a mesoscopic
equilibrium width. Within a cluster, individual particles tend to align in the direction
of gravity. 

The basic mechanism of instability and cluster formation
was described in a fundamental paper of Koch and Shaqfeh \cite{KS89}.
In Helzel and Tzavaras \cite{HT2015}, we recently derived a kinetic
model which describes the sedimentation process for dilute suspensions of rod-like
particles. By applying moment closure hypotheses and other approximations to
an associated moment system,  
we derived  macroscopic models for the evolution of the rod density
and compared the prediction of such macroscopic models to the original kinetic
model using numerical experiments. This is done in \cite{HT2015} for rectilinear
flows with the particles taking values on the sphere. 
Here, to simplify and explain our considerations we restrict to the simpler
case of shear flows and for particles with  orientations  restricted to take values on the plane.
While the derivations in \cite{HT2015} are often quite technical, 
the restriction to the simpler situation provides a useful and technically simple setting in order to
understand the underlying  ideas.  In addition, as explained in the present work,
the form of the derived macroscopic equations is identical in both cases apart from the values of numerical constants
that capture the effect of dimensionality in the microstructure.
Therefore, we hope that this paper will make our results accessible and useful for a wider community interested in
the modelling of complex fluids.
  
The rest of the paper is organised as follows. In Section
\ref{section:2} we present the kinetic model from \cite{HT2015} and a
simplification, which is obtained for shear flows. We then present
the simplified model system, obtained  by restricting the
orientation of particles to move in a plane. In Section \ref{section:3} we
derive a nonlinear moment closure system which forms the basis for  all further 
considerations. In Section \ref{section:4}, a linearization of the moment
closure system is used to study the linear instability of shear flows.
We show that a nonzero Reynolds number provides a wavelength
selection mechanism.  An asymptotic analysis of the largest eigenvalue
around $Re=0$ explains this behavior.
In Section \ref{section:5}, we depart from a closure of the non-linear moment system
and use a quasi-dynamic approximation in order to derive an effective equation 
for the evolution of the macroscopic density.
The basic assumption behind this approximation is that the behavior of the
second order moments can be replaced by equilibrium relations in the
moment closure system. In Section \ref{section:6}, we present a different set of
macroscopic equations for the density  obtained via the  so-called
diffusive limit.
We observe that the diffusive approximation leads to a typical 
Keller-Segel  model, while the quasi-dynamic approximation leads to a 
flux-limited Keller-Segel type model.   
In Section \ref{sec:numerics} we present numerical results
comparing the diffusive approximation and the quasi-dynamic
approximation to the  full kinetic model. Although the idea of diffusive scalings to obtain macroscopic
equations is commonplace in kinetic theory, it has not been applied (to our knowledge) in
the sedimentation problem.  For reasons of comparison, we present in an appendix,  
a derivation of the hyperbolic and diffusive scaling equations for general rectilinear flows
where the direction of the rod-like particles takes values on the sphere.

\section{A kinetic model for the sedimentation of rod-like
  particles}\label{section:2}
We consider a kinetic model for the sedimentation of rod-like
particles, derived in Helzel and Tzavaras \cite{HT2015},
based on kinetic models for dilute suspensions of rod-like particles that are
described in Doi and Edwards \cite{book:DE86}. 

In non-dimensional form, the kinetic model reads
\begin{equation}\label{eqn:system_nondim}
\begin{split}
\partial_t f  + \nabla_{\vec{x}} \cdot \left(\vec{u} f \right)
& +\nabla_{\vec{n}} \cdot \left(P_{\vec{n}^\bot} \nabla_{\vec{x}} \vec{u} \vec{n} f \right) - \nabla_{\vec{x}} \cdot \left(D(\vec{n})
\vec{e}_3 f\right)\\
&  =  D_r \Delta_{\vec{n}}f + \gamma \nabla_{\vec{x}} \cdot \left(
  D(\vec{n}) \nabla_{\vec{x}} f \right)\\
\sigma  & = \int_{S^{d-1}} \left( d \vec{n} \otimes \vec{n} - I \right) f\,  d \vec{n}\\
Re \left(\partial_{t} \vec{u} + \left( \vec{u} \cdot \nabla_{\vec{x}} \right) \vec{u}\right) & = \Delta_{\vec{x}} \vec{u} - \nabla_{\vec{x}} p
+ \delta \gamma \nabla_{\vec{x}} \cdot \sigma + \delta
\left( \brho - \int_{S^{d-1}} f \, d \vec{n} \right) \vec{e}_3\\
\nabla_{\vec{x}} \cdot \vec{u} & = 0.
\end{split}
\end{equation}
Here $f = f(t,\vec{x},\vec{n})$ is a distribution function of
particles at time $t$, space $\vec{x} \in \mathbb{R}^d$ and
orientation $\vec{n} \in S^{d-1}$, $\vec{u}=\vec{u}(t,\vec{x})$ is the
velocity of the solvent, $p=p(t,\vec{x})$ is the pressure and
$\sigma=\sigma(t,\vec{x})$ is the polymeric stress acting on the
fluid. Gradient, divergence and
Laplacian on the sphere are denoted by $\nabla_{\vec{n}}$,
$\nabla_{\vec{n}} \cdot$ and $\Delta_{\vec{n}}$, while gradient and
divergence in the macroscopic space are denoted by $\nabla_{\vec{x}}$
and $\nabla_{\vec{x}} \cdot$.
The term  
$$
P_{\vec{n}^\bot}
\nabla_{\vec{x}} \vec{u} \vec{n} = \nabla_{\vec{x}} \vec{u} \vec{n} -
\left( \nabla_{\vec{x}} \vec{u} \vec{n} \cdot \vec{n} \right) \vec{n}
$$
is the projection of the vector $\nabla_{\vec{x}} \vec{u} \vec{n}$
onto the tangent space at $\vec{n}$ while $D(\vec{n}) = I + \vec{n} \otimes \vec{n}$. 
Finally, the unit vector $-\vec{e}_3$ points into the direction of gravity.
We note that $D_r$, $\delta$, $\gamma$ and $Re$ are non-dimensional numbers
and refer to \cite{HT2015} for their description.
Furthermore, the constant $\brho$  describes the average density of the suspended rods;  
it can be easily absorbed into the pressure, by modifying it to account for the hydrostatic pressure, 
but we prefer to put it as a factor that equilibrates the body force. 

The second and the fourth term on the left hand side of (\ref{eqn:system_nondim})$_1$  model transport 
of the center of mass of the particles due to the macroscopic flow velocity and due  to gravity, respectively. 
The third term  on the left hand side models the effect of 
rotation of the axis of the microstructure due to a macroscopic velocity gradient. The 
terms on the right hand side of the first equation in
(\ref{eqn:system_nondim}) describe rotational and translational
diffusion respectively.
Equation (\ref{eqn:system_nondim})$_2$ defines the stress tensor emerging from the microstructure
while (\ref{eqn:system_nondim})$_3$ and (\ref{eqn:system_nondim})$_4$ describe the macroscopic flow 
of the solvent.

In this paper we mainly restrict our considerations to the study of
shear flows, i.e.\ we consider the case $\vec{u} = (0, 0,
w(t,x))^T$. The director $\vec{n}$ may in general take values in
$S^2$, which means that the rod-like particles are allowed to move out
of the plane described by the direction of shear and the direction of gravity. 
Furthermore, we will often consider the case $\gamma = 0$. 

For shear flow,  the system (\ref{eqn:system_nondim}) simplifies to an
equation of the 
form
\begin{equation}\label{eqn:full-shear}
\begin{split}
& \partial_t f(t,x,\vec{n}) + \nabla_{\vec{n}} \cdot 
\left( P_{\vec{n}^\bot} (0,0,n_1  w_x)^T f \right) - \partial_x (n_1 n_2 f) = D_r \Delta_{\vec{n}} f \\
& Re \, \partial_t w(t,x) = \partial_{xx} w + \delta \left( \brho - \int_{S^2}
  f\, d\vec{n} \right).
\end{split}
\end{equation}

An even simpler system is obtained, if we restrict the director
to only take values in the $(x,z)$ plane, i.e.\ the plane spanned by the direction of
shear and the direction of gravity. (While this restriction is not natural, it turns out to
be much simpler to analyze and in retrospect to provide the same form of macroscopic equations.)
In this situation we have
$\vec{n} \in S^1$, i.e.\ $\vec{n} =  (\cos \theta, \sin \theta)^T$
with $\theta \in [0,2\pi)$.  The angle $\theta$ is measured counter clockwise
from the positive $x$-axis. 
In this simplified situation, the model for shear flow can be written in the form
\begin{equation}\label{eqn:simple-shear}
\begin{split}
& \partial_t f(t,x,\theta) + \partial_\theta \left( w_x \cos^2 \theta f \right)
- \partial_x \left( \sin \theta \cos \theta f \right)
= D_r \partial_{\theta \theta} f \\
& Re \, \partial_t w(t,x) = \partial_{xx} w + \delta \left( \brho - \int_0^{2\pi} f
  \, d\theta \right).
\end{split}
\end{equation}

\section{Nonlinear moment closure}\label{section:3}

In this section we consider \eqref{eqn:simple-shear} and proceed to describe the evolution of
a system of moments. The macroscopic density is
$$
\rho(t,\vec{x}) = \int_{S^1} f(t, \vec{x}, \vec{n}) \, d\vec{n}.
$$
We use as basis for the moments the eigenfunctions of the Laplace-Beltrami operator $\del_{\theta \theta}$ on
the circle $S^1$.
These are the functions $1$, and $\cos n \theta$, $\sin n \theta$, $n =1, 2, 3, ...$. Since the rods are identical
under the reflection $\theta \to - \theta$ only the even eigenfunctions will play a role.

First we derive a nonlinear system of equations for the zero-th
order moment $\rho$ and the second order moments $s$ and $c$, 
 defined via the relations
\begin{equation*}
\begin{split}
\rho(t,x) & := \int_0^{2\pi} f(t,x,\theta) \, d\theta \\
c(t,x) & := \frac{1}{2} \int_0^{2 \pi} f(t,x,\theta) \cos(2\theta) \,
d\theta \\
s(t,x) & := \frac{1}{2} \int_0^{2 \pi} f(t,x,\theta) \sin(2 \theta) \, d\theta .
\end{split}
\end{equation*}
We will close the system by neglecting the moments of order higher than $2$. This closure is based on
the premise that higher moments will experience faster decay, as they correspond to a larger eigenvalue of the 
Laplace-Beltrami operator. The validity of this hypothesis will be tested numerically.

In our derivation we  consider the different terms of the first equation in
(\ref{eqn:simple-shear}) separately, using the notation
\begin{equation}\label{eqn:f-shear-2}
\begin{split}
\partial_t f & = \underbrace{- \partial_\theta \left( w_x \cos^2 \theta \, f\right)}_{[1]}
+ \underbrace{\partial_x \left( \sin \theta \cos \theta f \right)}_{[2]} +
\underbrace{D_r \partial_{\theta \theta} f}_{[3]}. 
\end{split}
\end{equation}
The evolution equation for $\rho$ is obtained by integrating
(\ref{eqn:f-shear-2}) over $S^1$. Note that there is no contribution
from $[1]$ and $[3]$, and 
\begin{equation*}
\begin{split}
\partial_t \rho & = \int_0^{2 \pi} \partial_t f \, d\theta \\
& = \int_0^{2 \theta} \partial_x \left( \sin \theta \cos \theta f
\right) \, d \theta \\
& = \partial_x s.
\end{split}
\end{equation*}
To obtain the evolution equation for $c$, we compute
\begin{equation*}
\begin{split}
\partial_t c & = \frac{1}{2} \int_0^{2 \pi} \cos(2\theta) \partial_t f
\, d\theta.
\end{split}
\end{equation*}
We separately consider the different contributions from the right hand
side of (\ref{eqn:f-shear-2}).

\noindent
contribution from $[1]$:
\begin{equation*}
\begin{split}
 - \frac{1}{2} \int_0^{2 \pi} \cos (2\theta) \partial_\theta \left((
  w_x \cos^2 \theta f \right) \, d\theta 
& = - w_x \int_0^{2\pi} \cos(2\theta) \partial_\theta \left(
  \frac{1}{2} (1 + \cos 2\theta ) f \right) \, d\theta \\
& = - w_x \int_0^{2 \pi} \sin(2\theta)  \frac{1}{2} (1 + \cos(2\theta)
f \, d\theta \\
& = - w_x s - \frac{1}{4} w_x \int_0^{2 \pi} \sin(4 \theta) f \,
d\theta  
\end{split}
\end{equation*}

\noindent
contribution from $[2]$:
\begin{equation*}
\begin{split}
\frac{1}{2} \int_0^{2\pi} \cos(2\theta) \partial_x \left(
  \sin \theta \cos \theta \, f \right) \, d\theta & 
= \frac{1}{4} \int_0^{2\pi} \cos(2\theta) \sin(2\theta) \partial_x f
\, d\theta \\
& = \frac{1}{8} \int_0^{2\pi} \sin(4\theta) \partial_x f \, d\theta
\end{split}
\end{equation*}

\noindent
contribution from $[3]$:
\begin{equation*}
\begin{split}
\frac{1}{2} \int_0^{2 \pi} \cos(2\theta) D_r \partial_{\theta \theta}
f \, d\theta & = D_r \int_0^{2\pi} \sin(2\theta) \partial_\theta f \,
d\theta \\
& = -2 D_r \int_0^{2\pi} \cos(2\theta) f \, d\theta \\
& = -4 D_r c
\end{split}
\end{equation*}

Now we neglect higher order moments, i.e.\ terms that involve
integrals of the form \linebreak
$\int_0^{2\pi} \sin(4 \theta) f \, d\theta$ and
$\int_0^{2\pi} \cos(4\theta) f \, d\theta$.
Under this assumption, the evolution equation for $c$ has the form
$$
\partial_t c = - w_x s - 4 D_r c.
$$
Similarly we can derive an evolution equation for $s$. The complete
nonlinear moment closure system reads 
\begin{equation}\label{eqn:qd-moments}
\begin{split}
\partial_t \rho & = \partial_x s \\
\partial_t c & = - w_x s - 4 D_r c \\
\partial_t s & = \frac{1}{8} \partial_x \rho + w_x c + \frac{1}{4} w_x
\rho - 4 D_r s\\
Re \partial_t w & = \partial_{xx} w + \delta \left( \brho - \rho \right).
\end{split}
\end{equation}

\section{Linear stability theory}\label{section:4}
We now consider the linear stability of the shear flow problem and
give an asymptotic expansion of the unstable eigenvalue of the linear
system in the Reynolds number $Re$. 

We linearize the moment closure system (\ref{eqn:qd-moments}) around
the state $w=0$ and $\rho = 1$. To simplify the notation we set $\gamma =
0$, $D_r = 1$ and consider the system
\begin{equation}\label{eqn:linear-moments}
\begin{split}
\partial_t \rho & = \partial_x s \\
\partial_t c & = - 4 c \\
\partial_t s & = \frac{1}{8} \partial_x \rho + \frac{1}{4} w_x - 4 s
\\
Re \partial_t w & = \partial_{xx} w - \delta \rho.
\end{split}
\end{equation}
Fourier transformation of the first three equations of
(\ref{eqn:linear-moments}) leads to the system
\begin{equation}
\begin{split}
\partial_t \hat{\rho}(\xi) & = i \xi \hat{s}(\xi) \\
\partial_t \hat{c}(\xi) & = - 4 \hat{c}(\xi) \\
\partial_t \hat{s}(\xi) & = \frac{1}{8} i \xi \hat{\rho}(\xi) +
\frac{1}{4} i \xi \hat{w}(\xi) - 4 \hat{s}(\xi).
\end{split}
\end{equation}
First we consider the case  $Re = 0$. In this case Fourier
transformation  of the last equation of (\ref{eqn:linear-moments})
leads to
\begin{equation}
\xi^2 \hat{w}(\xi) = - \delta  \hat{\rho}(\xi).
\end{equation} 

Thus, we obtain the linear
system (dropping the hats)  
\begin{equation}\label{eqn:A1}
\partial_t \left( \begin{array}{c}
\rho \\ c \\ s\end{array}\right) = \left( \begin{array}{ccc}
0 & 0 & i \xi \\
0 & -4 & 0\\
\frac{1}{8} i\xi - \frac{i \delta}{4 \xi} & 0 & -4\end{array}\right)
\left( \begin{array}{c}
\rho \\ c \\ s\end{array}\right).
\end{equation}
The matrix on the right hand side of (\ref{eqn:A1}) has the
eigenvalues
\begin{equation*}
\left\{ -4, -2 - \frac{1}{4} \sqrt{64 + 4 \delta - 2 \xi^2},
  -2+\frac{1}{4}\sqrt{64+4\delta-2\xi^2}\right\}
\end{equation*}
The last eigenvalue, which we denote by $\lambda_0$, is larger than
zero provided that $\delta >0$ and $\xi^2$ is small enough. Thus, the
linear moment closure system coupled with the Stokes equation is most
unstable for waves with wave number $\xi \rightarrow 0$. 
The eigenvector corresponding to the eigenvalue $\lambda_0$ has the
form
$$
\vec{x}_0 = \left( \frac{2 i \xi}{2 \delta - \xi^2} 
\left(8+\sqrt{64+4\delta-2\xi^2}\right), 0, 1\right)^T.
$$

Now we consider the case $Re >0$, for which the linearized coupled 
system can be expressed in the form 
\begin{equation}\label{eqn:asymp_4}
\begin{split}
\left( \begin{array}{c}
\dot{\vec{x}}\\\dot{y}\end{array} \right) = \left( \begin{array}{c c}
A & B \\ \frac{1}{Re} D & \frac{1}{Re} C \end{array} \right) 
\left( \begin{array}{c}
\vec{x} \\ y \end{array} \right).
\end{split}
\end{equation}
with $\vec{x}=(\rho, c, s)^T$, $y=w$ and
\begin{equation*}
\begin{split}
A & = \left( \begin{array}{ccc}
0 & 0 & i\xi\\
0 & -4 & 0\\
\frac{1}{8} i\xi & 0 & -4\end{array}\right), \, B =
\left( \begin{array}{c}
0\\
0\\
\frac{1}{4} i \xi\end{array}\right), 
D  = (\begin{array}{ccc}-\delta & 0 & 0\end{array}), \, 
C = -\xi^2
\end{split}
\end{equation*}
Our goal is to give an asymptotic expansion for eigenvalues of the matrix
arising on the right hand side of (\ref{eqn:asymp_4}), which is valid for 
small values of $Re$. In particular we wish to understand how the
eigenvalue $\lambda_0$ changes with $Re$. We make the ansatz
\begin{equation}
\begin{split}
A \vec{x}_\varepsilon + B y_\varepsilon & = \lambda_\varepsilon \vec{x}_\varepsilon\\
C y_\varepsilon + D \vec{x}_\varepsilon & = \varepsilon \lambda_\varepsilon y_\varepsilon,
\end{split}
\end{equation} 
with
\begin{equation*}
\begin{split}
\lambda_\varepsilon & = \lambda_0 + \varepsilon {\lambda}_1 + \ldots\\
\vec{x}_\varepsilon & = \vec{x}_0 + \varepsilon \vec{x}_1 + \ldots\\
y_\varepsilon & = y_0 + \varepsilon y_1 + \ldots.
\end{split}
\end{equation*}
We obtain

${\cal O}(1)$: (as considered above)
\begin{equation}\label{eqn:O1}
\begin{split}
A \vec{x}_0 + B y_0 & = \lambda_0 \vec{x}_0\\
C y_0 + D \vec{x}_0 & = 0
\end{split}
\end{equation}

${\cal O}(\varepsilon)$:
\begin{equation}\label{eqn:asymp_5}
\begin{split}
A \vec{x}_1 + B y_1 & = \lambda_0 \vec{x}_1 + \lambda_1 \vec{x}_0\\
C y_1 + D \vec{x}_1 & = \lambda_0 y_0,
\end{split}
\end{equation}
which we rewrite into the form
\begin{equation}\label{eqn:asymp_6}
\left( \begin{array}{cc}
A - \lambda_0 I & B\\
D & C\end{array} \right) \left( \begin{array}{c}
\vec{x}_1 \\ y_1 \end{array} \right) = \left( \begin{array}{c}
\lambda_1 \vec{x}_0\\
\lambda_0 y_0 \end{array} \right).
\end{equation}
The left null space of the matrix on the left hand side of (\ref{eqn:asymp_6})
is given by
\begin{equation}
\vec{u}^T = \left( -8-\sqrt{64+4\delta-2\xi^2}, 0, -4 i \xi,  1\right).
\end{equation}
We multiply both sides of Equation (\ref{eqn:asymp_6}) from the left 
 with $\vec{u}^T$  and obtain
\begin{equation}\label{eqn:asymp_7}
0 = \vec{u}^T \cdot \left( \begin{array}{c}
\lambda_1\vec{x}_0 \\
\lambda_0 y_0 \end{array} \right).
\end{equation}
From the second equation of (\ref{eqn:O1}) we obtain
$$
y_0 = - C^{-1} D \vec{x}_0 = \frac{-2 i \delta}{\xi}
\frac{(8+\sqrt{64-2\xi^2+4\delta})}{2\delta - \xi^2}.
$$
Using the expressions for $\lambda_0$, $\vec{x}_0$, $y_0$ and $\vec{u}$
in Equation (\ref{eqn:asymp_7}), we can now calculate $\lambda_1$,
which has the form
\begin{equation}
\lambda_1 = - \frac{\delta}{\frac{2 \xi^2}{(2 \delta - \xi^2)} (8 +
  \sqrt{64+4\delta-2\xi^2})^2 + 4 \xi^2}.
\end{equation}

\begin{figure}[htbp!]
(a)\includegraphics[width=0.4\textwidth]{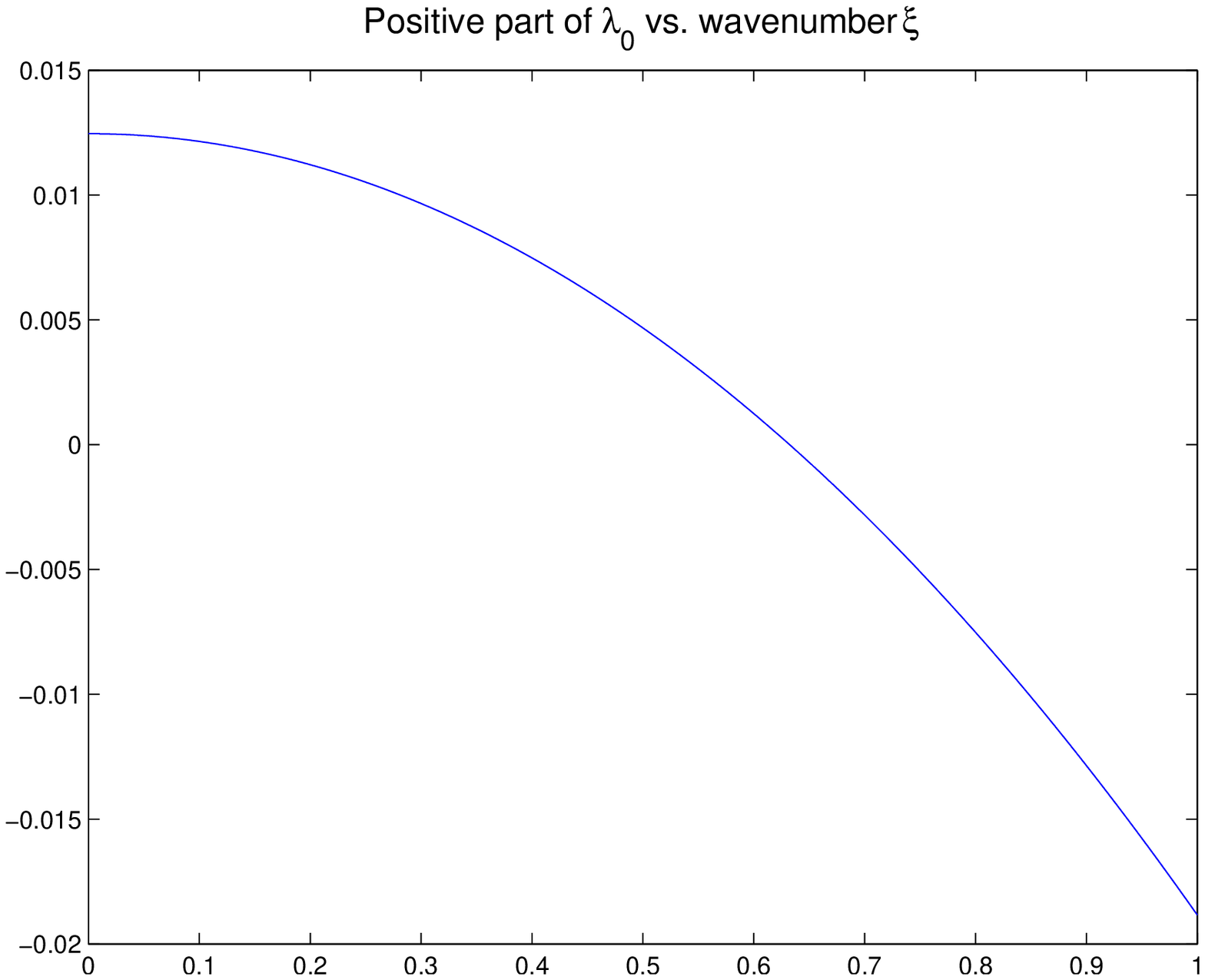}\hfill
(b)\includegraphics[width=0.4\textwidth]{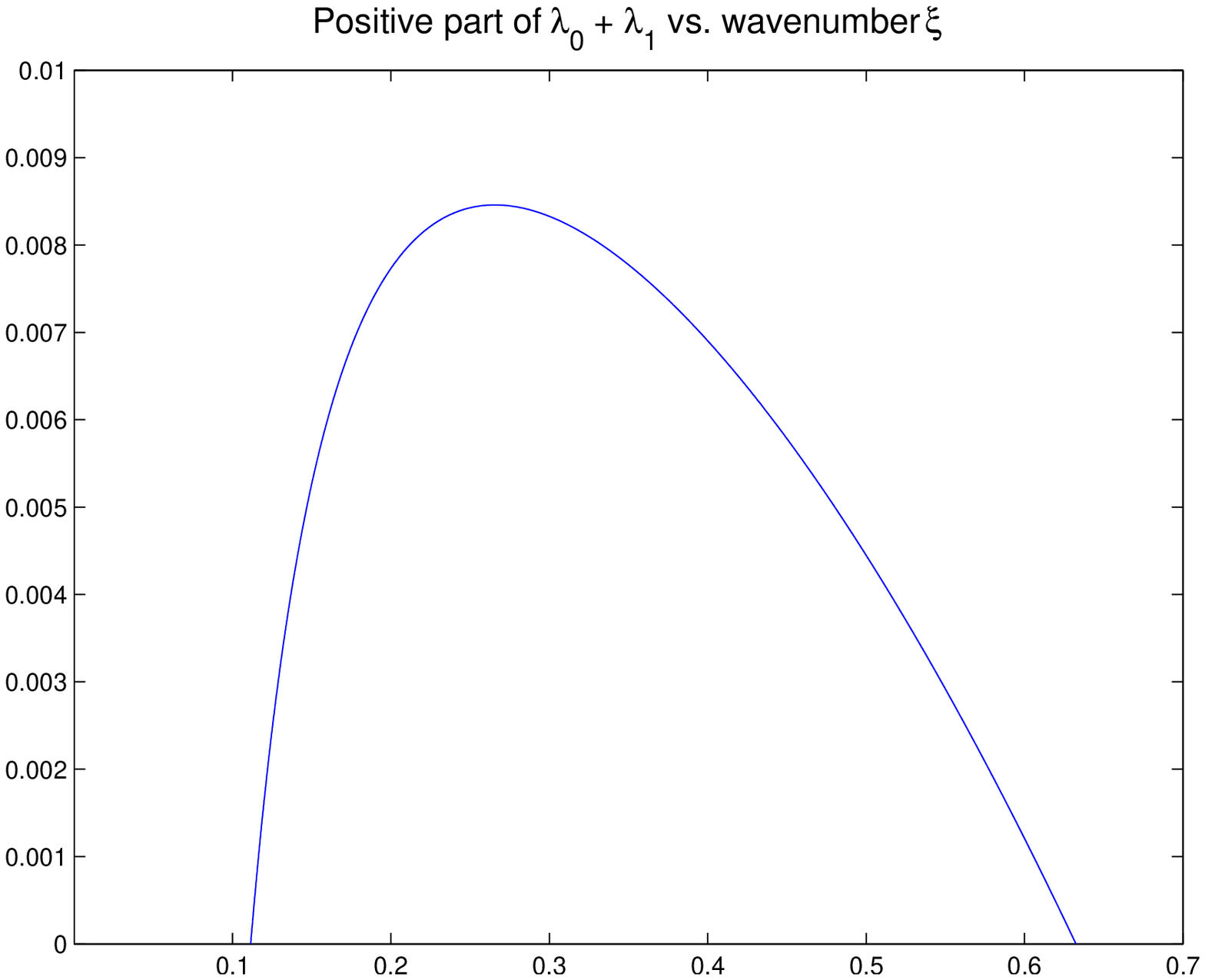}
\caption{\label{fig:asymp1}Positive part of the (a) $\lambda_0$ and
  (b) $\lambda_0 + \lambda_1$ vs.\ wave number $\xi$ for $\delta = 0.2$.}
\end{figure}

In Figure \ref{fig:asymp1}(a) we plot the eigenvalue $\lambda_0$ as a
function of the wave number $\xi$. This eigenvalue describes the
linear stability behavior of the system for $Re = 0$. The longest
possible waves are most unstable and there is no wave length selection.
In Figure \ref{fig:asymp1}(b), we plot
$\lambda_0 + Re \lambda_1$ vs.\ the wave number $\xi$ for $\varepsilon
= Re =1$ and $\delta = 0.2$. A non-zero Reynolds number provides a
wave length selection mechanism.
The curve attains its maximum for a wave number $\xi \approx 0.27$.
Thus, the linear theory predicts the wave length of the most unstable wave 
to be approximately $2\pi / 0.27 \approx 23.27$. On an interval of
length 100, we expect the formation of 4-5 clusters. This is in
agreement with our numerical result shown in Section
\ref{sec:numerics}, Figure \ref{fig:numerics-2}.

\section{The quasi-dynamic approximation}\label{section:5}
In \cite{HT2015}, we derived various systems of evolution equations describing 
the macroscopic behavior of  the system
(\ref{eqn:full-shear}) for intermediate and long times.  
We considered rectilinear flows in the direction of gravity with the
rod orientations taking values 
on the sphere $S^2$ and derived the so called quasi-dynamic approximation ({\it cf} \cite{HT2015}).
When one restricts to a shear flow in the direction of gravity
$$
\vec{u} = (0, 0, w(t,x))^T \, , \qquad f = f(t, x, \vec{n}), \quad \vec{n} \in S^2
$$
and set $D_r =1$  we obtain the system consisting of an advection-diffusion equation 
coupled to a diffusion equation,
\begin{equation}\label{eqn:qd-shear}
\begin{split}
&\partial_t \rho(t,x) = \frac{1}{30}\partial_x \left[ \frac{42^2}{42^2 +
    46 w_x^2} \left(  \rho w_x + \frac{1}{3}\partial_x \rho \right)
\right]\\
& Re \, \partial_t w(t,x) = \partial_{xx} w + \delta \left( \brho - \rho \right).
\end{split}
\end{equation}
This model belongs to the class of flux-limited Keller-Segel systems and enjoyes gradient-flow structure (see \cite{HT2015}).
The derivation of \eqref{eqn:qd-shear}   is quite technical due to the  complex  form of the moment equations
(and their closures) reflecting the structure of harmonic polynomials in dimension $d > 2$.

To shed some light on the derivation of \eqref{eqn:qd-shear}, we consider the 
simplified shear flow model (\ref{eqn:simple-shear}), where now the rigid-rods are constrained to move in-plane, 
$\vec{n} \in S^1$, and present a derivation of the quasi-dynamic
approximation.
The idea behind the quasi-dynamic approximation is that the transient
dynamics of the second order moments $c$ and $s$ is replaced by its
equilibrium response, i.e.\ by
\begin{equation}\label{eqn:qd-gls}
\begin{split}
4 D_r c + w_x s & = 0 \\
-\frac{1}{8} \partial_x \rho  - \frac{1}{4} w_x \rho - w_x c + 4 D_r s
& = 0.
\end{split}
\end{equation}
System (\ref{eqn:qd-gls}) can be solved for $s$, to obtain
\begin{equation}\label{eqn:wq-s}
s = \frac{D_r w_x \rho + \frac{1}{2} D_r \partial_x \rho}{16 D_r^2 + w_x^2}.
\end{equation}
Inserting (\ref{eqn:wq-s}) into the first equation of
(\ref{eqn:qd-moments}), we obtain the quasi-dynamic approximation
\begin{equation}
\begin{split}
\partial_t \rho & = \partial_x \left( \frac{D_r w_x}{16 D_r^2 + w_x^2}
  \rho \right) + \frac{1}{2} D_r \partial_x \left( \frac{1}{16 D_r^2 +
    w_x^2} \partial_x \rho \right) \\
Re \partial_t w & = \partial_{xx} w + \delta \left( \brho-\rho \right).
\end{split}
\end{equation}
For the special case $D_r = 1$ it gives
\begin{equation}\label{eqn:dq-simple}
\begin{split}
\partial_t \rho & = \partial_x \left( \frac{1}{16 + w_x^2} \left(
   w_x \rho + \frac{1}{2} \partial_x \rho \right) \right) \\
Re \partial_t w & = \partial_{xx} w + \delta \left( \brho -\rho \right).
\end{split}
\end{equation}
A comparison of (\ref{eqn:dq-simple}) with (\ref{eqn:qd-shear}) shows
that we obtain the same general structure of a flux-limited
Keller-Segel model. The simplification, which restricts the director to
$S^1$, just leads to different constants. 
 
\section{The diffusive scaling}\label{section:6}

The diffusive scaling provides another approach for obtaining a
macroscopic evolution equation for $\rho$. 
Here we present a direct derivation of the diffusive scaling for the
simplified shear flow model (\ref{eqn:simple-shear}), where we assume
that the director $f$ only takes values on $S^1$. Again we restrict our
considerations to the case $D_r = 1$.

We consider (\ref{eqn:simple-shear}) and rescale the model 
in the diffusive
scale, i.e.
$$
x = \frac{1}{\delta} \hat{x}, \, t = \frac{1}{\delta^2} \hat{t}, \,
\vec{u} = \hat{\vec{u}}.
$$ 
The scaled equations (dropping the hats and for $D_r =1$) have the form
\begin{equation}\label{eqn:simple-shear-scaled}
\begin{split}
&\delta^2 \, \partial_t f(t,x,\theta) + \delta \, 
\partial_\theta \left( w_x \cos^2 \theta f \right)
-\delta  \partial_x \left( \sin \theta \cos \theta f \right)
=  \partial_{\theta \theta} f \\
& Re \, \delta^2 \, \partial_t w(t,x) = \delta^2 \, \partial_{xx} w + \delta \left( \brho - \int_0^{2\pi} f
  \, d\theta \right).
\end{split}
\end{equation}
Now we introduce the ansatz
\begin{equation*}
\begin{split}
f(t,x,\theta) & = \delta f_0 + \delta^2 f_1 + \ldots \\
\vec{u} & = \vec{u}_0 + \delta \vec{u}_1 + \ldots = \left( \begin{array}{c}
0 \\ w_0 \end{array} \right) + \delta \left( \begin{array}{c}
0 \\ w_1 \end{array} \right) + \ldots\\
\brho & = \delta \brho_0 
\end{split}
\end{equation*}
and obtain the relations
\begin{align}
&O(\delta)  \quad 
&&\partial_{\theta \theta} f_0 = 0 
\label{as1-simple}
\\
&O(\delta^2)  \qquad 
&& \partial_{\theta} \left( w_x \cos^2 \theta f_0 \right) - \partial_x
\left( \sin \theta \cos \theta f_0 \right) = \partial_{\theta \theta}
f_1 \label{as2-simple}\\
& && Re \partial_t w_0 = \partial_{xx} w_0 + \left( m - \int_0^{2 \pi}
  f_0 d\theta \right) 
\nonumber
\\
&O(\delta^3) \qquad
&& \partial_t f_0 + \partial_\theta \left( w_x \cos^2 \theta f_1
\right) - \partial_x \left( \sin \theta \cos \theta f_1 \right)
= \partial_{\theta \theta} f_2.
\label{as3-simple}
\end{align}
The relation (\ref{as1-simple}) implies that $f_0$ depends at most
linearly on $\theta$. As a function on $S^1$, $f$ is periodic. Thus,
$f_0$ must be constant in $\theta$ and
\begin{equation*}
\rho_0(t, x) = \int_0^{2\pi} f_0(t, x) d\theta = 2 \pi f_0(t, x).
\end{equation*}
Now we integrate equation (\ref{as3-simple}) over $S^1$. The
second and the last term vanish and we obtain
\begin{equation*}\label{as4-simple}
\partial_t \rho_0 - \partial_x \int_0^{2 \pi} \cos \theta \sin \theta
f_1 \, d \theta = 0.
\end{equation*}
Using integration by parts, we arrive at
\begin{equation*}
\begin{split}
\partial_t \rho_0 & = - \frac{1}{4} \partial_x \left(
  \int_0^{2\pi} \partial_{\theta \theta} \left( \cos \theta \sin
    \theta \right) f_1 \, d\theta \right)\\
& = - \frac{1}{4} \partial_x \left( \int_0^{2\pi} \cos \theta \sin
  \theta \, \partial_{\theta \theta} f_1 \, d\theta \right).
\end{split}
\end{equation*}
Using (\ref{as2-simple}), we 
replace $\partial_{\theta \theta} f_1$ by terms which depend on $f_0$ and
obtain an evolution equation for $\rho_0$.
\begin{equation*}
\begin{split}
\partial_t \rho_0 & = -\frac{1}{4} \partial_x \left( 
\int_0^{2\pi} \cos \theta \sin \theta \left[ \partial_\theta \left(
    w_x \cos^2 \theta f_0 \right) - \partial_x \left( \sin \theta \cos
    \theta f_0 \right) \right] \, d \theta \right) \\
& =  \frac{1}{4 \pi} \, \partial_x \left( w_x \rho_0
  \int_0^{2\pi} \cos^2 \theta \sin^2 \theta \, d \theta \right) 
+ \frac{1}{8 \pi} \partial_x \left( 
(\partial_x \rho_0 ) \int_0^{2 \pi} \cos^2 \theta \sin^2 \theta \,
d\theta \right) \\
& = \frac{1}{16} \partial_x \left( w_x \rho_0 \right) +
\frac{1}{32} \partial_{xx} \rho_0
\end{split}
\end{equation*}
Thus, in the diffusive limit, the dynamics of the simplified shear
flow problem (\ref{eqn:simple-shear}) is described by the system
\begin{equation}\label{eqn:diff-simple}
\begin{split}
\partial_t \rho & = \frac{1}{16} \partial_x \left( w_x \rho  +
\frac{1}{2} \partial_{x} \rho \right ) \\
Re \, \partial_t w & = \partial_{xx} w + \left( \brho - \rho \right).
\end{split}
\end{equation}

In Appendix \ref{app:diff},
we derive the diffusive scaling for the kinetic model
(\ref{eqn:system_nondim}) in the case of
rectilinear flow for rigid rods that move in $S^2$. For shear flow and in the special case $D_r = 1$,
$\gamma = 0$, the diffusive scaling of (\ref{eqn:full-shear}) leads to the model equation
\begin{equation}\label{eqn:diff-limit-shear}
\begin{split}
\partial_t \rho & = \frac{1}{30} \partial_x \left( w_x \rho +
  \frac{1}{3} \partial_x \rho \right) \\
Re \partial_t w & = \partial_{xx} w + \left( \brho - \rho \right).
\end{split}
\end{equation} 
Again we obtain a Keller-Segel type model. 
In contrast to the quasi-dynamic approximation, the diffusive scaling does not
provide flux limiting.

\section{Numerical simulations}\label{sec:numerics}
\begin{figure}[htbp!]
\includegraphics[width=0.4\textwidth]{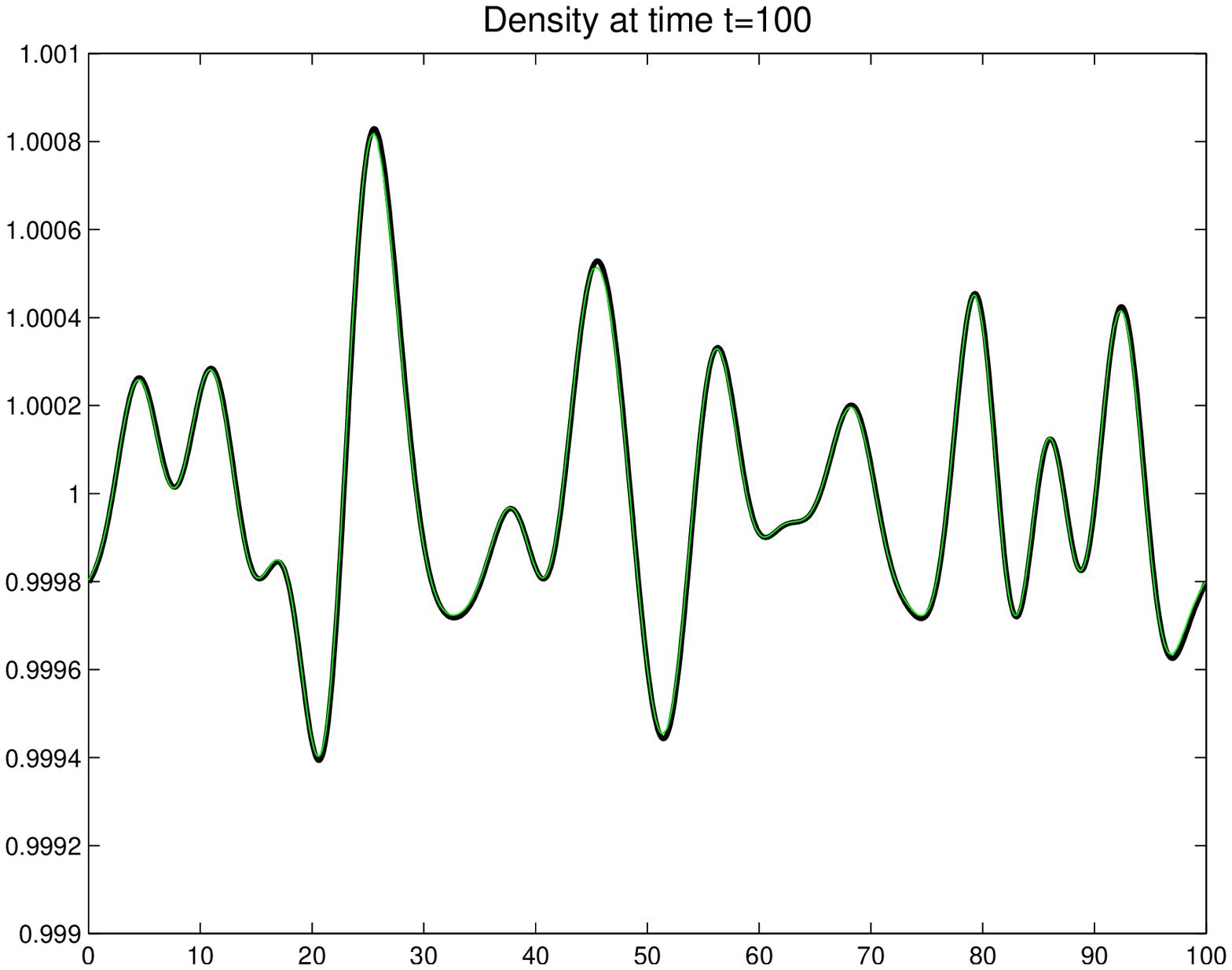}\hfill
\includegraphics[width=0.4\textwidth]{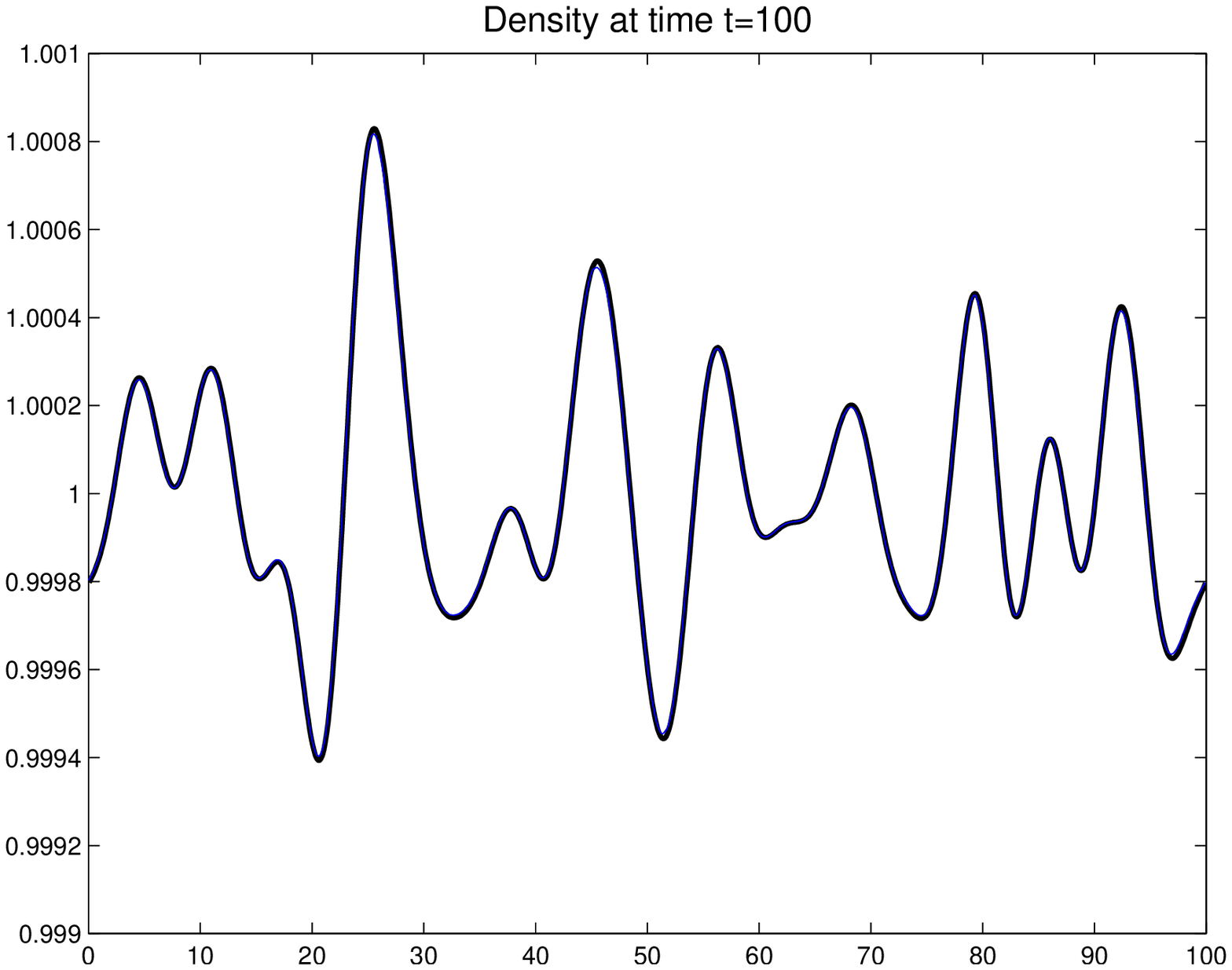}

\includegraphics[width=0.4\textwidth]{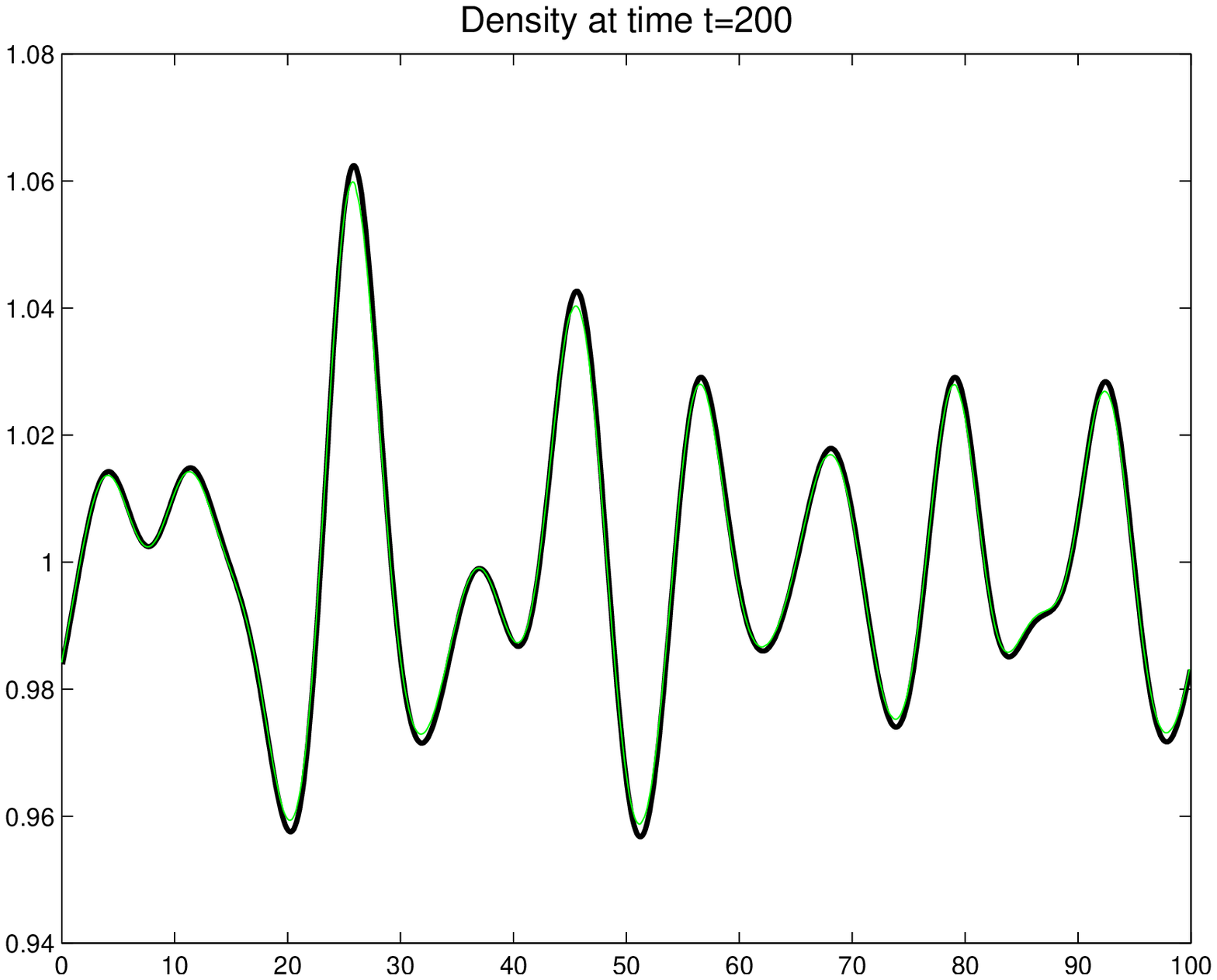}\hfill
\includegraphics[width=0.4\textwidth]{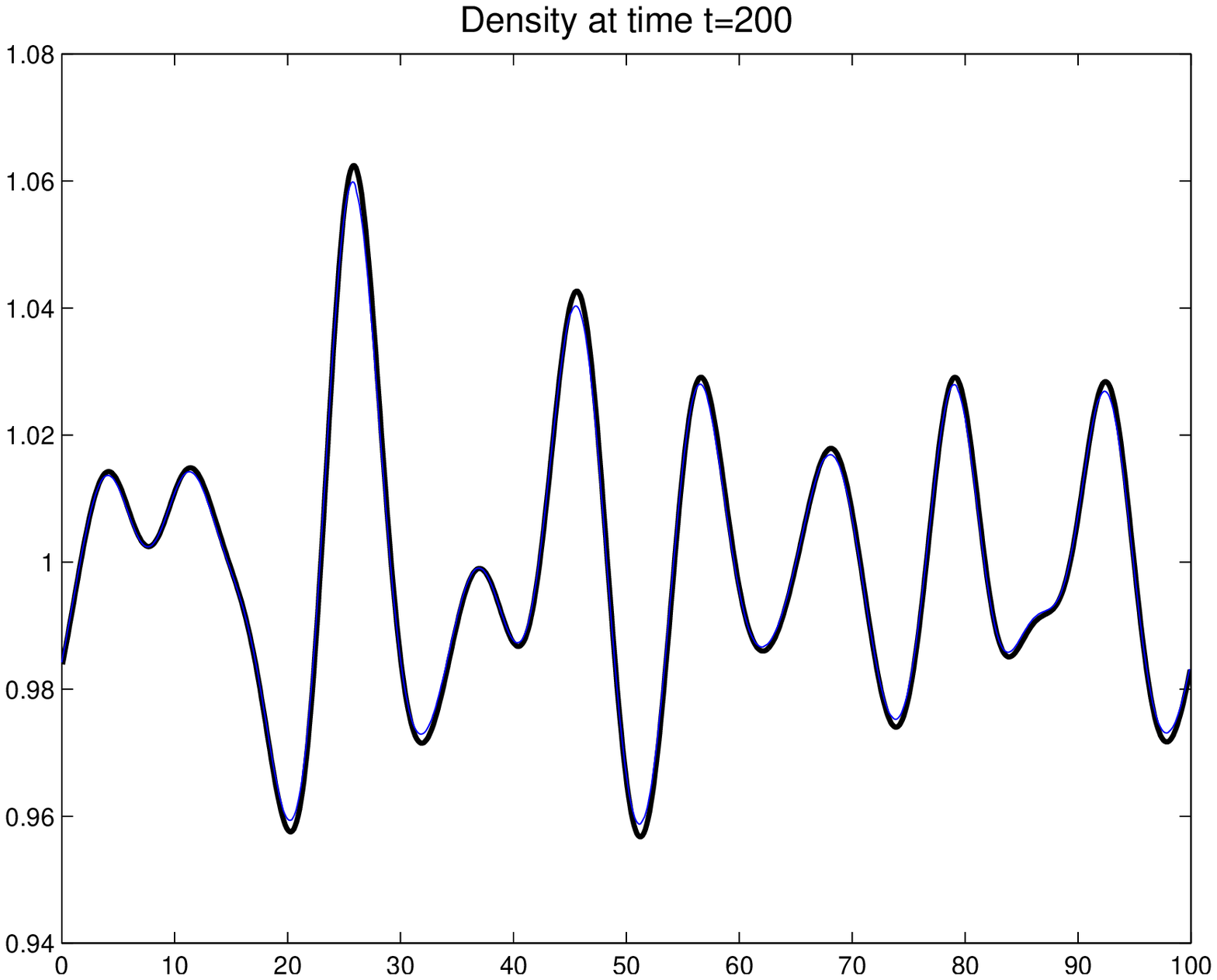}

\includegraphics[width=0.4\textwidth]{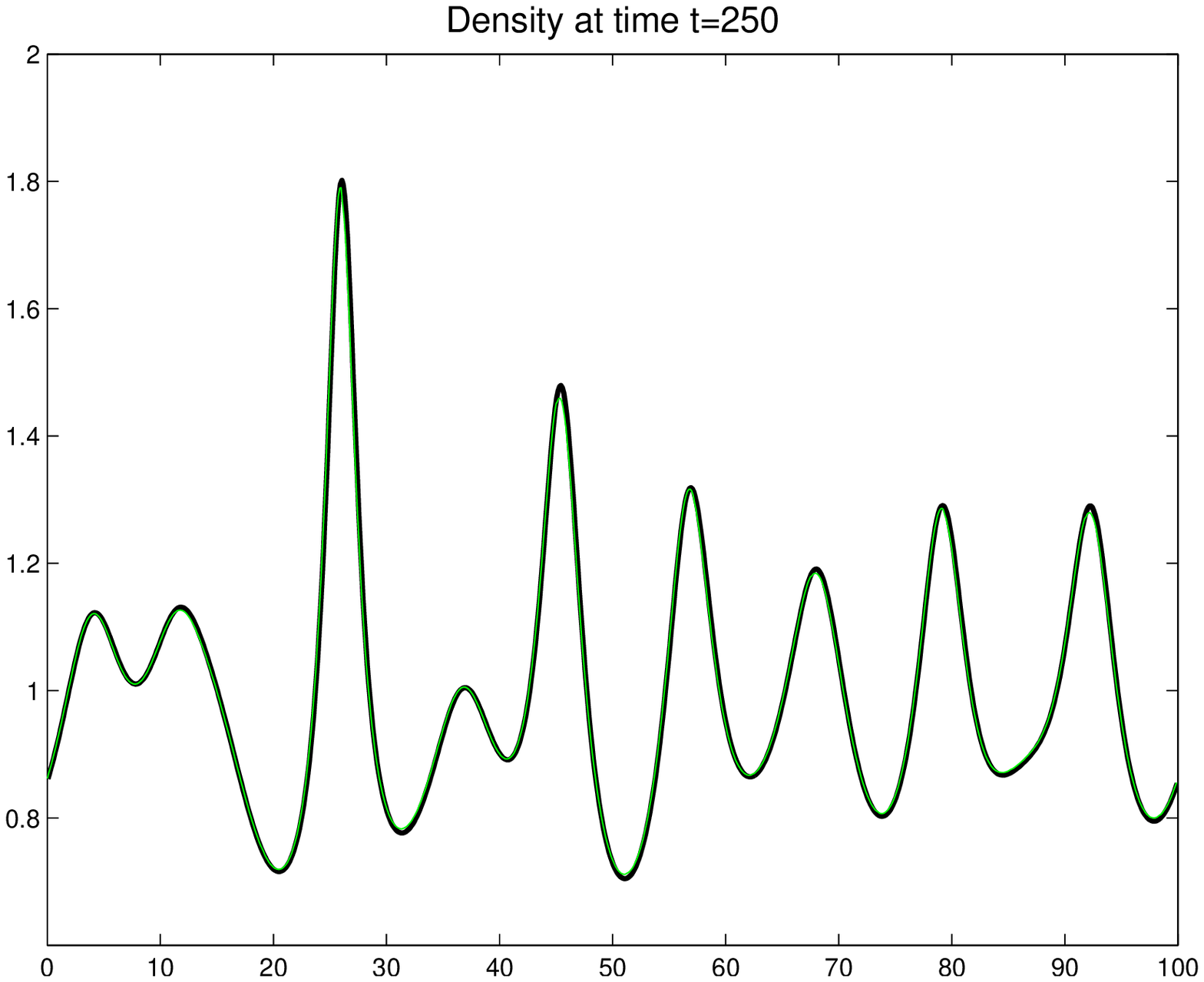}\hfill
\includegraphics[width=0.4\textwidth]{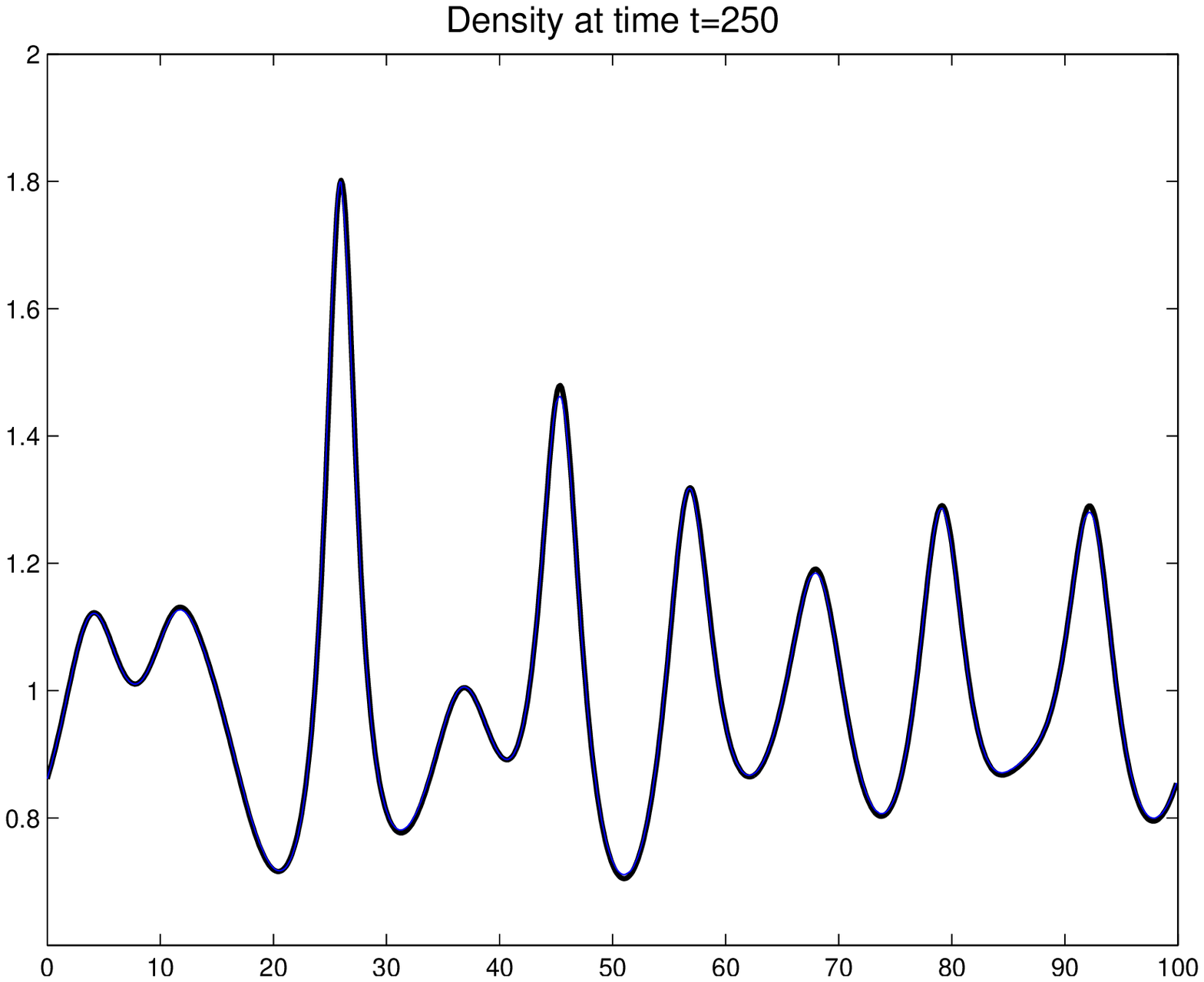}

\includegraphics[width=0.4\textwidth]{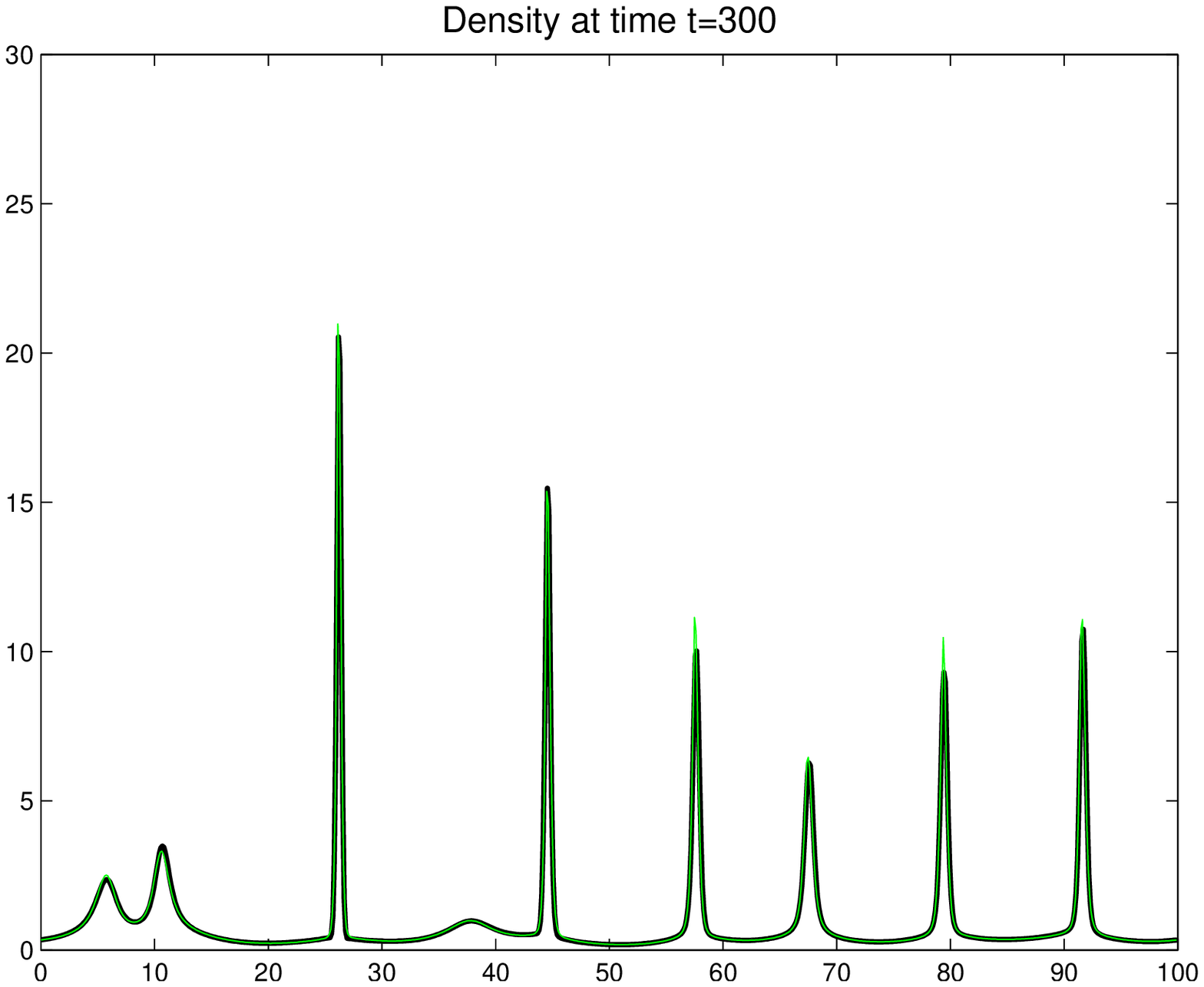}\hfill
\includegraphics[width=0.4\textwidth]{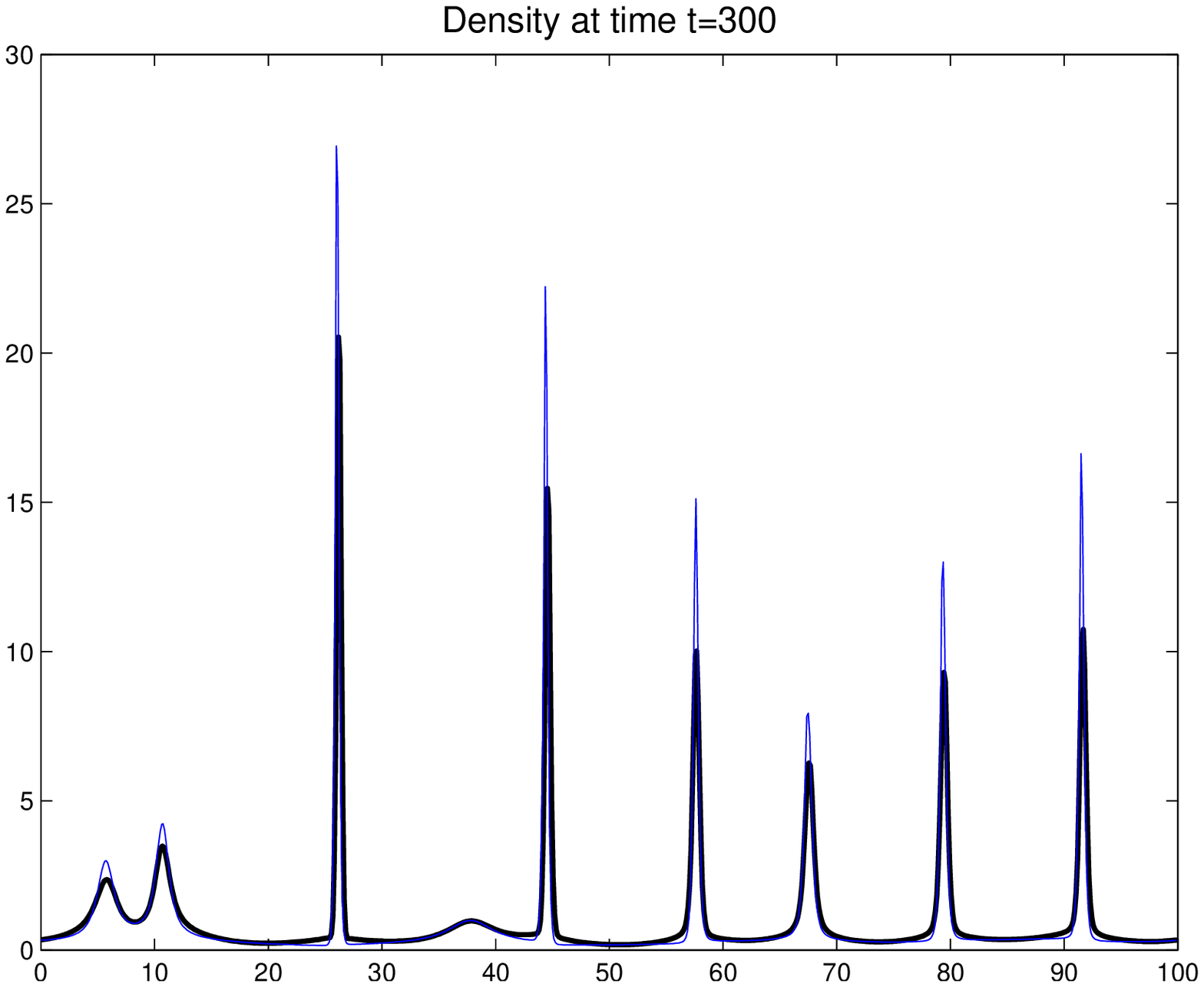}
\caption{\label{fig:numerics-1}
Left plots: Comparison of the quasi-dynamic approximation (green line) with
  the full model (black line). Right plots: Comparison of the
  diffusive scaling (blue line) with the full model (black line).}
\end{figure}

In this section we show numerical simulations for shear flow, which compare the
simplified shear flow model (\ref{eqn:simple-shear}) with the quasi-dynamic
approximation (\ref{eqn:qd-shear}) and the diffusive
scaling (\ref{eqn:diff-simple}).
In Figure \ref{fig:numerics-1} we show results of numerical
simulations using the parameter values $D_r = \delta = Re = 1$. The
initial values are set to be  
\begin{equation} \label{eqn:numerics-iv}
\begin{split}
\rho(x_k,0) & = 1 + 10^{-4} \left( \epsilon(x_k) - \frac{1}{2}
\right) \\
w(x_k,0) & = 0,
\end{split}
\end{equation}
where $\epsilon(x_k)$ is a random number between $0$ and $1$. 
We impose the periodicity condition on an interval of length $100$.
For our test simulations we used $800$ grid cells in space, thus
$k=1,\ldots,800$. In the simulation of the full model, $S^1$
is discretized with $200$ grid cells. We observe the formation of
clusters with higher particle density.
Both the solutions predicted by the quasi-dynamic approximation as
well as the solutions predicted by the diffusive limit
compare very well with the solution structure obtained by the full
model. Only at a very late times, some differences can be observed and
the quasi-dynamic approximation leads to slightly more accurate results.

\begin{figure}[htbp!]
\includegraphics[width=0.4\textwidth]{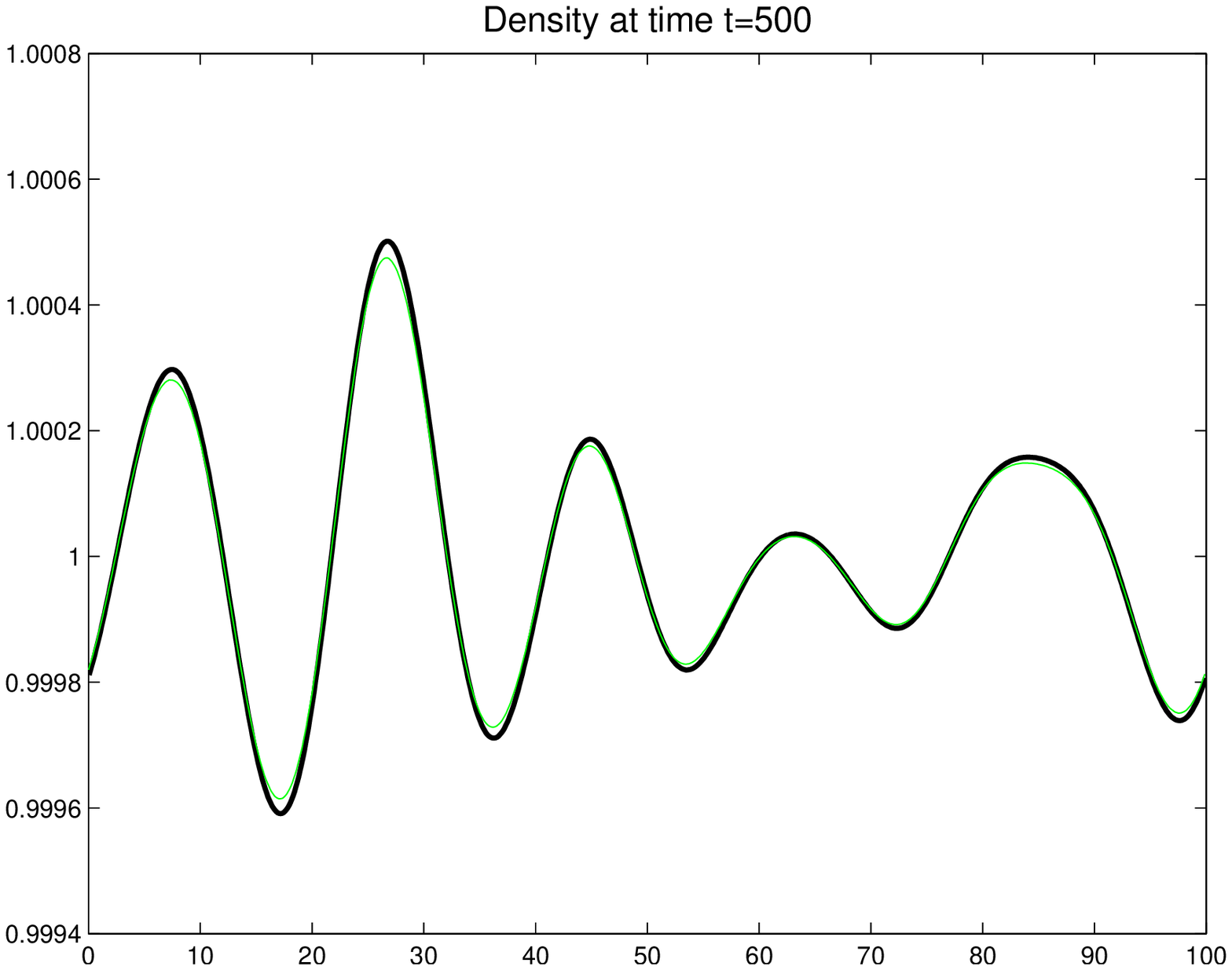}\hfill
\includegraphics[width=0.4\textwidth]{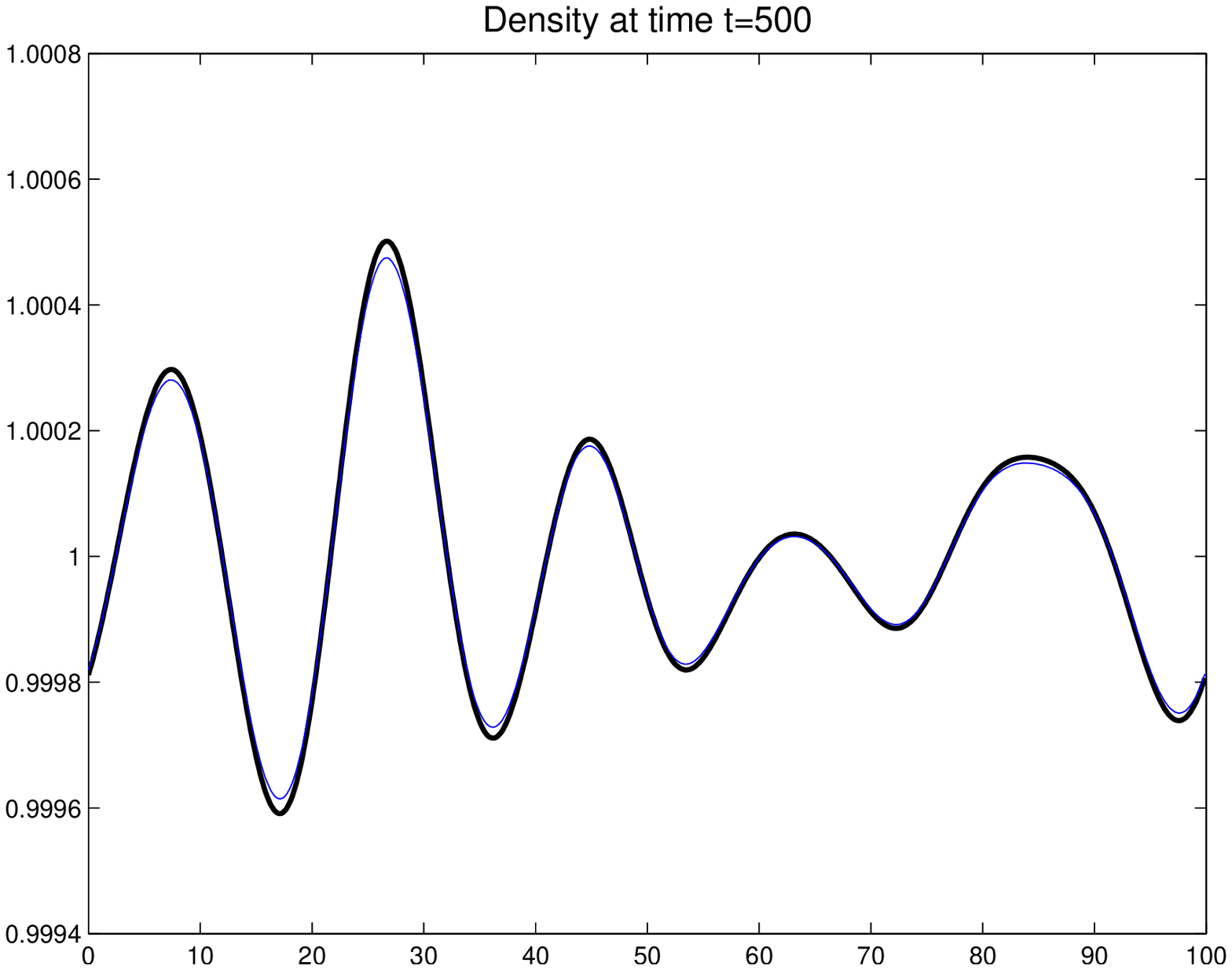}

\includegraphics[width=0.4\textwidth]{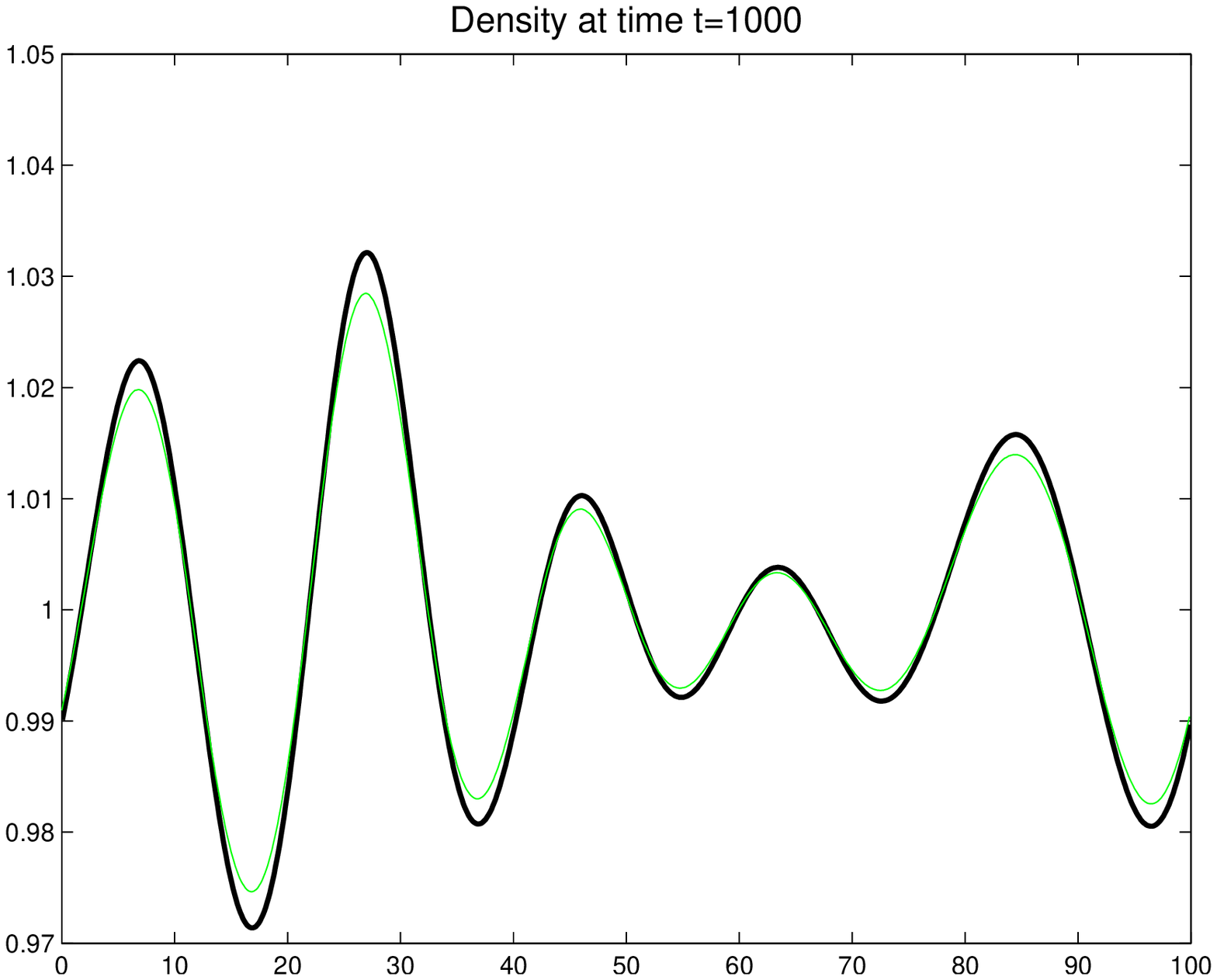}\hfill
\includegraphics[width=0.4\textwidth]{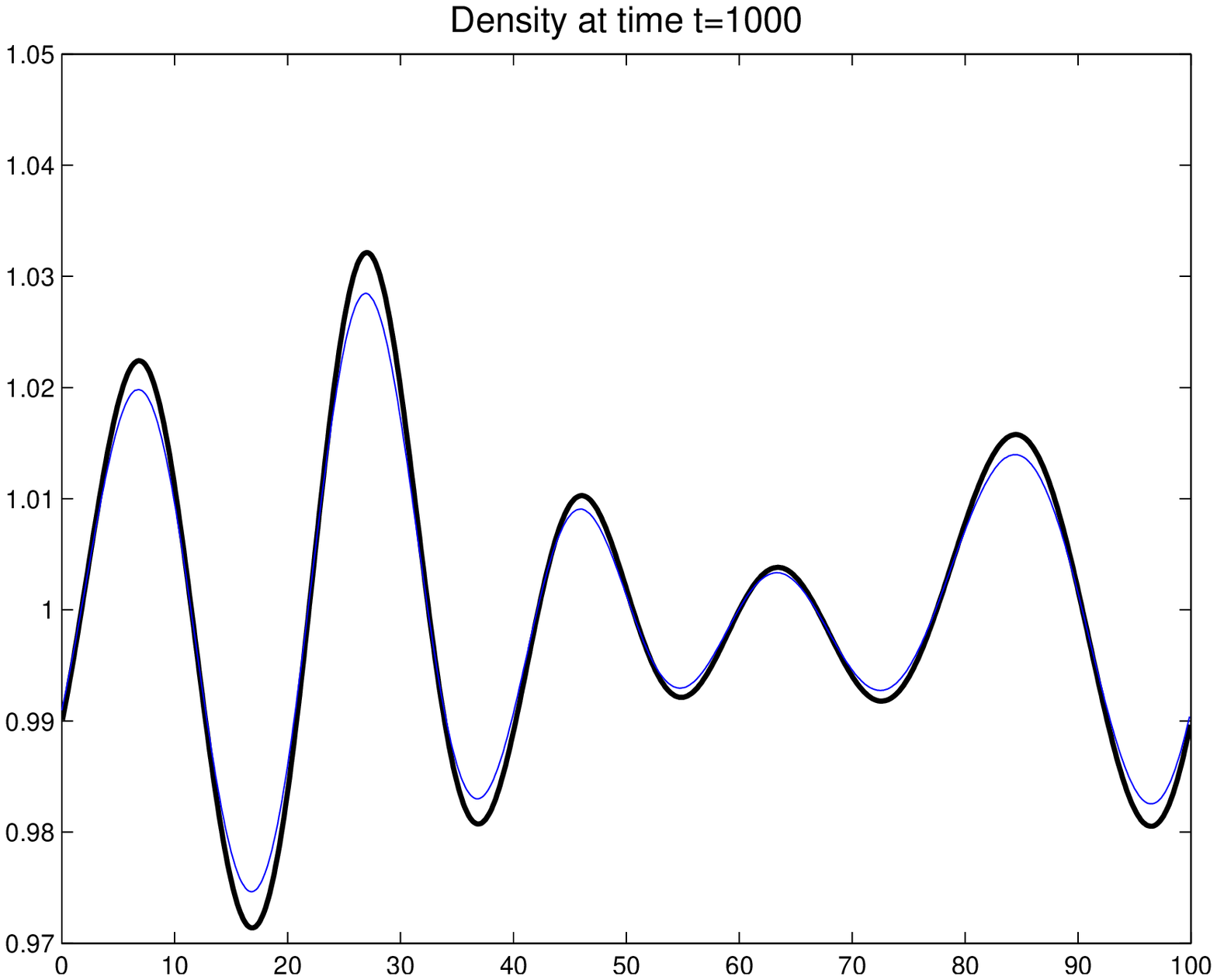}

\includegraphics[width=0.4\textwidth]{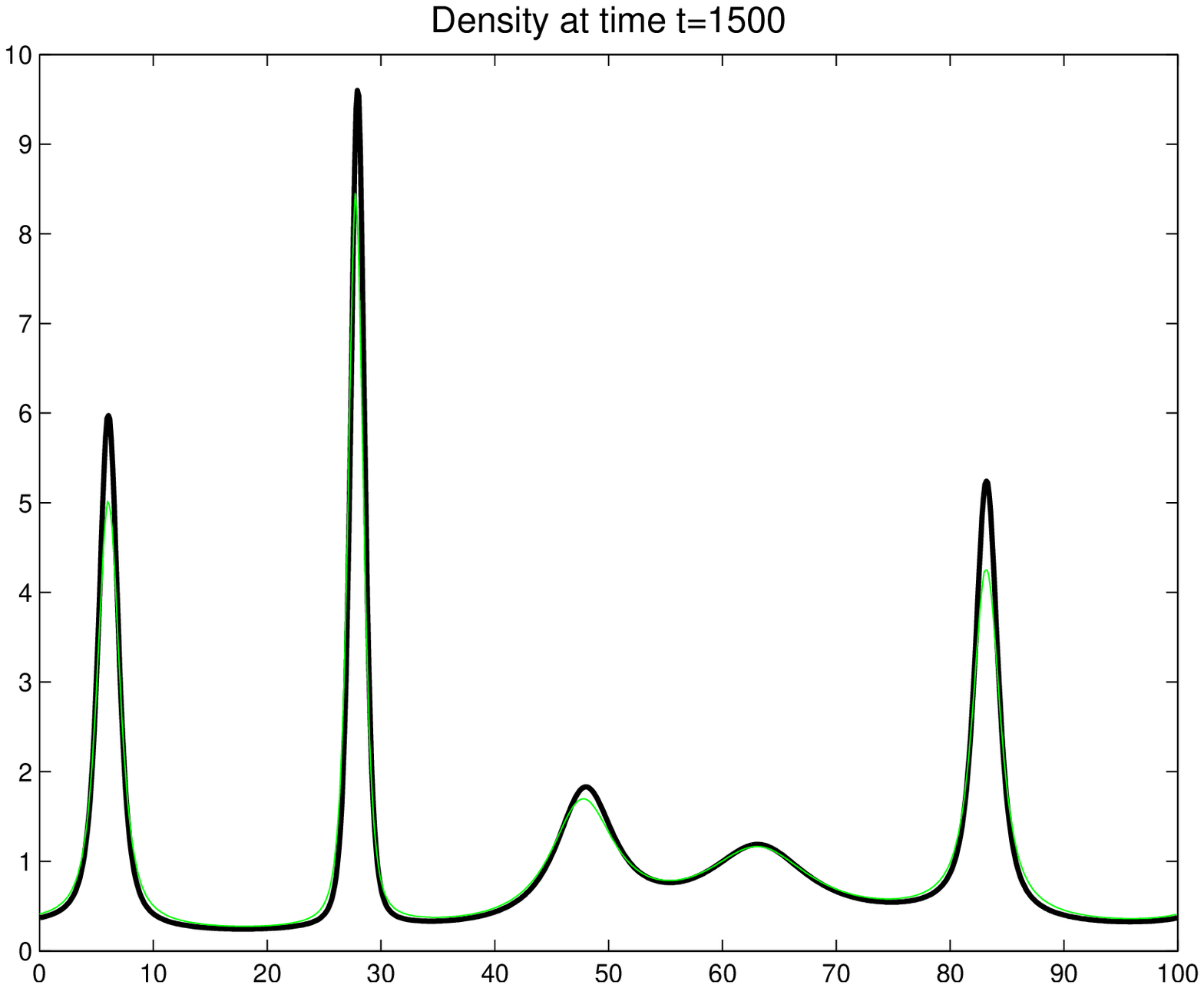}\hfill
\includegraphics[width=0.4\textwidth]{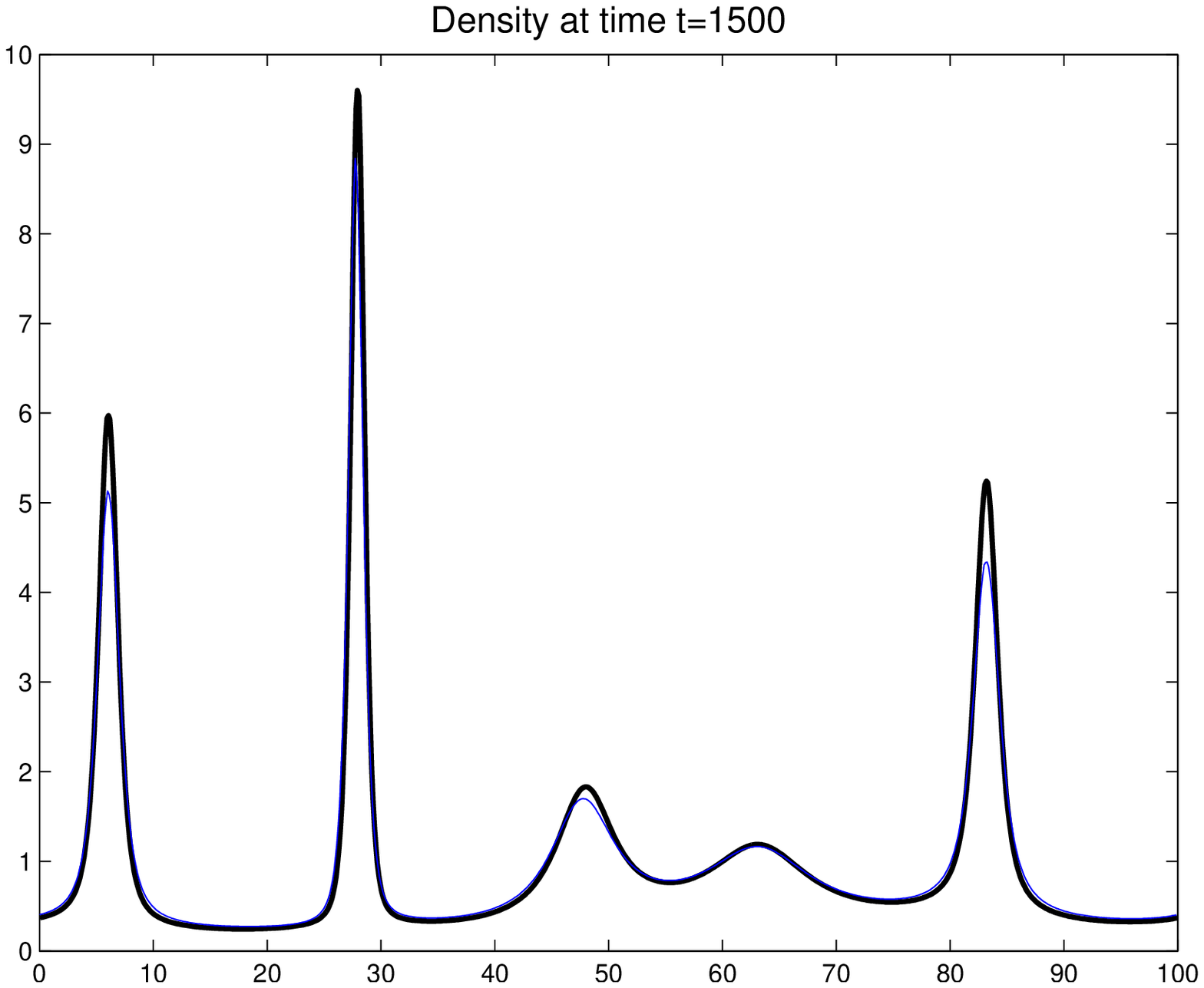}

\includegraphics[width=0.4\textwidth]{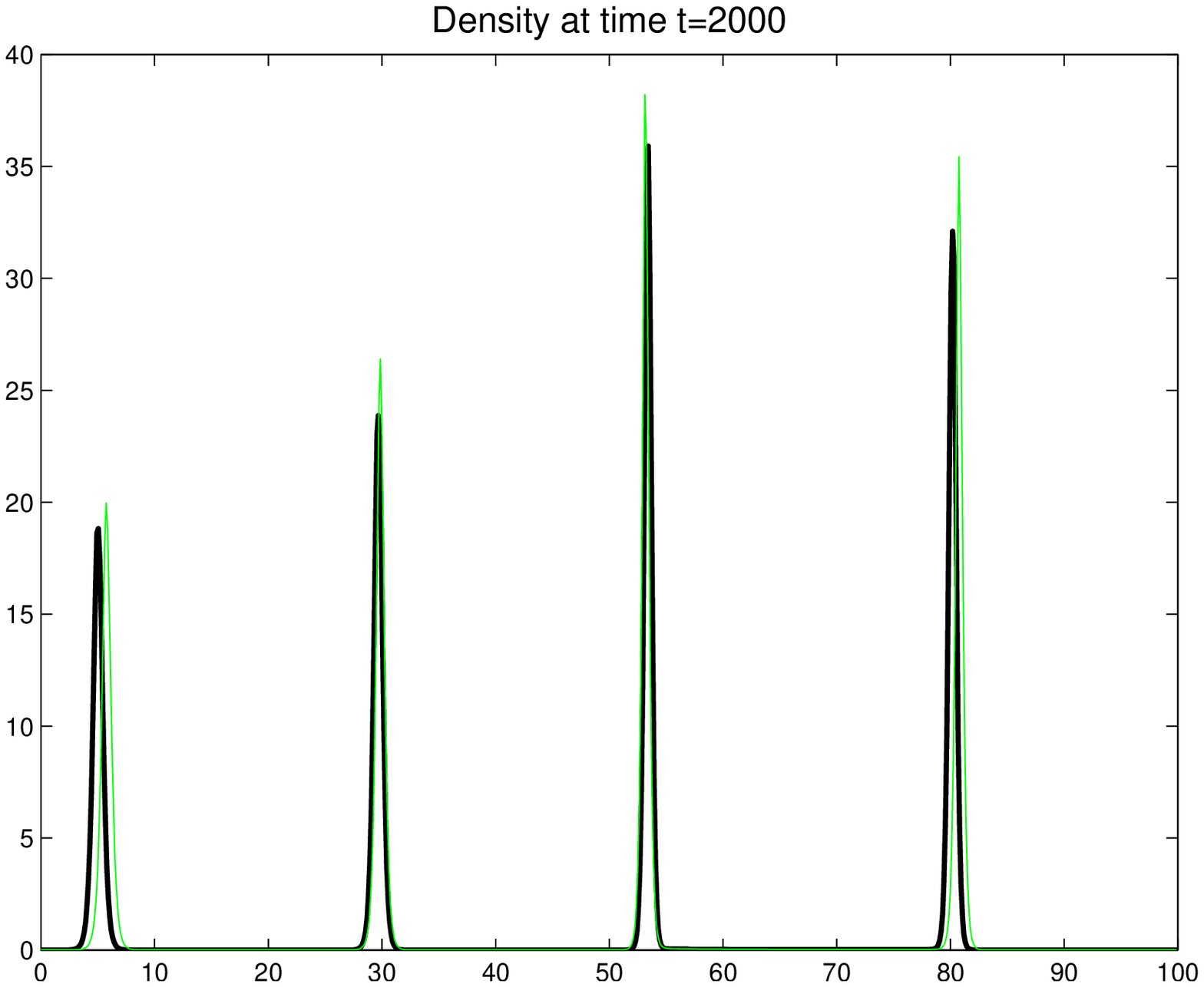}\hfill
\includegraphics[width=0.4\textwidth]{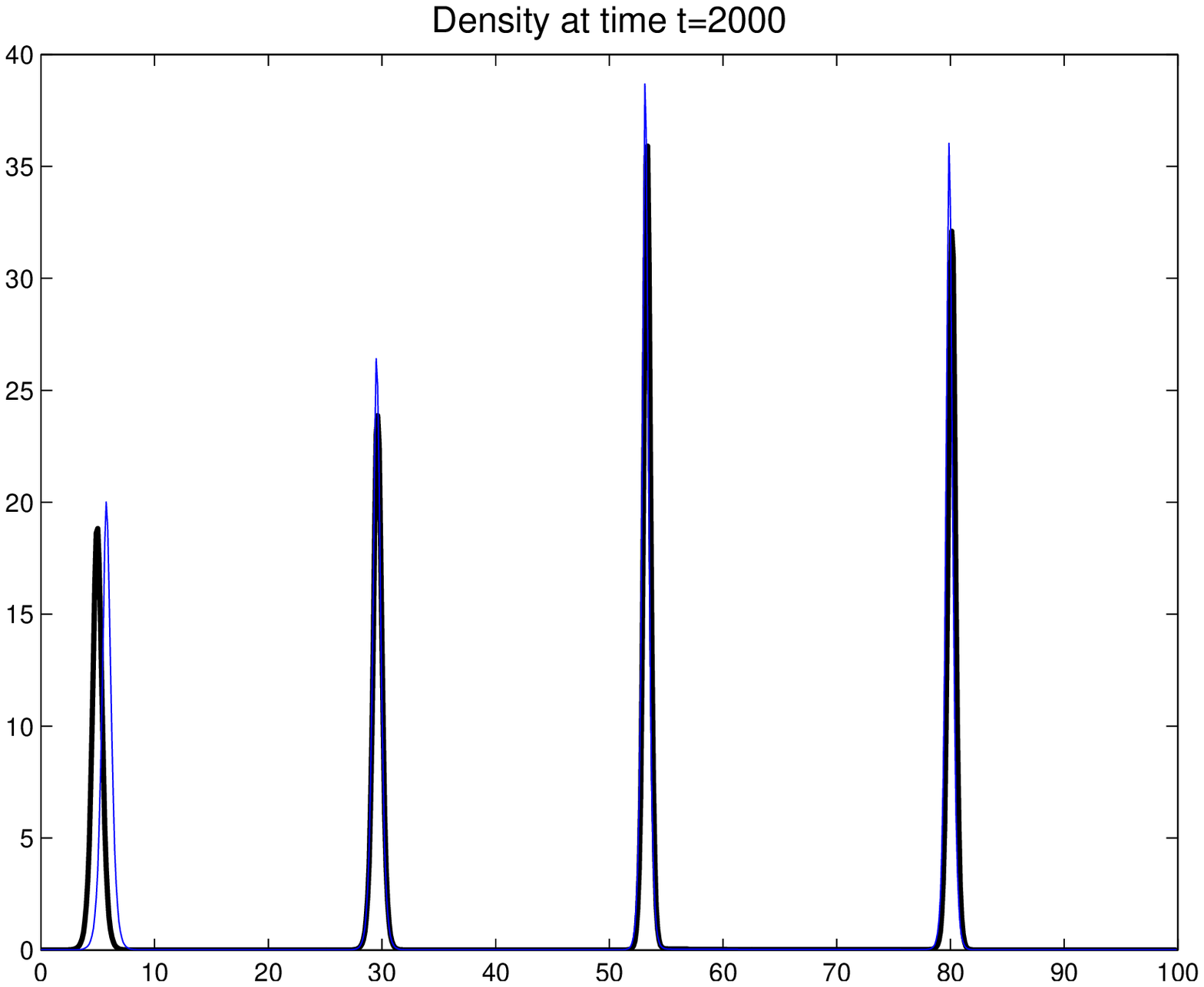}
\caption{\label{fig:numerics-2}
Left plots: Comparison of the quasi-dynamic approximation (green line) with
  the full model (black line). Right plots: Comparison of the
  diffusive scaling (blue line) with the full model (black line).
The simulations correspond to  $\delta = 0.2$.}
\end{figure}

Note that the model equations obtained by the quasi-dynamic approximation
contain the same non-dimensional parameters as the full model. For the
diffusive limit this is not the case, since the parameter $\delta$
does no longer appear in (\ref{eqn:diff-simple}). In Figure
\ref{fig:numerics-2} 
we show simulations
comparing the quasi-dynamic approximation with the full model for  
$\delta = 0.2$, using the same initial conditions as above.

In order to simulate this problem  with the diffusive model, we impose
periodic boundary conditions on a domain of length $\delta 100 = 20$
and consider numerical approximations for $t \le \delta^2 2000=80$.
Furthermore, we set initial values as described by
(\ref{eqn:numerics-iv}) but multiplied by $1/\delta$. Finally we used
$\brho = 1/\delta$ in equation (\ref{eqn:diff-simple}). 
To compare the numerical solution of the diffusive limit
system with the solution of the full model, we map the numerical
solution to the interval $[0,100]$ and multiply with $\delta$. The
results are shown in Figure \ref{fig:numerics-2} on the right hand side.

\begin{figure}[htbp!]
\includegraphics[width=0.4\textwidth]{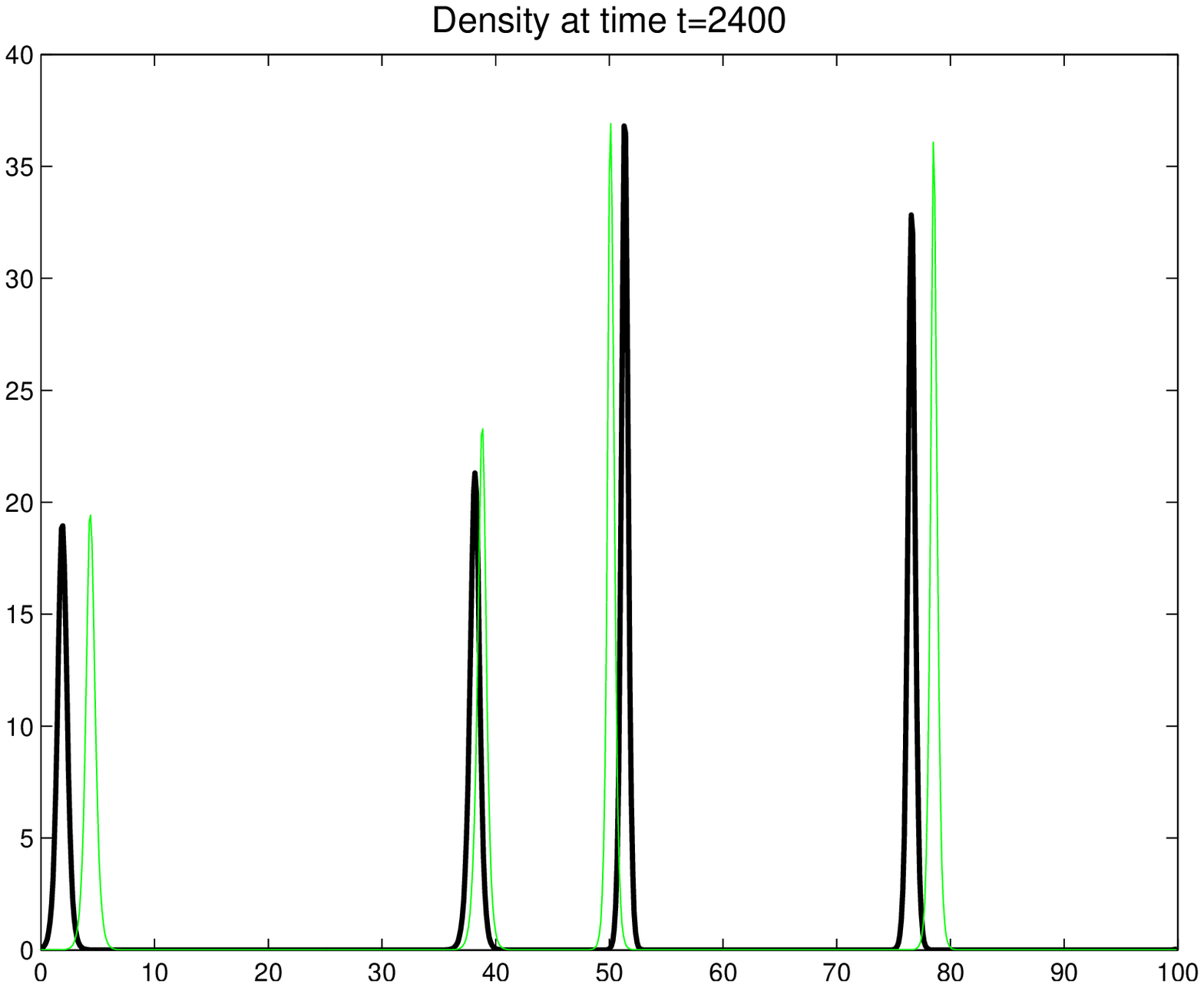}\hfill
\includegraphics[width=0.4\textwidth]{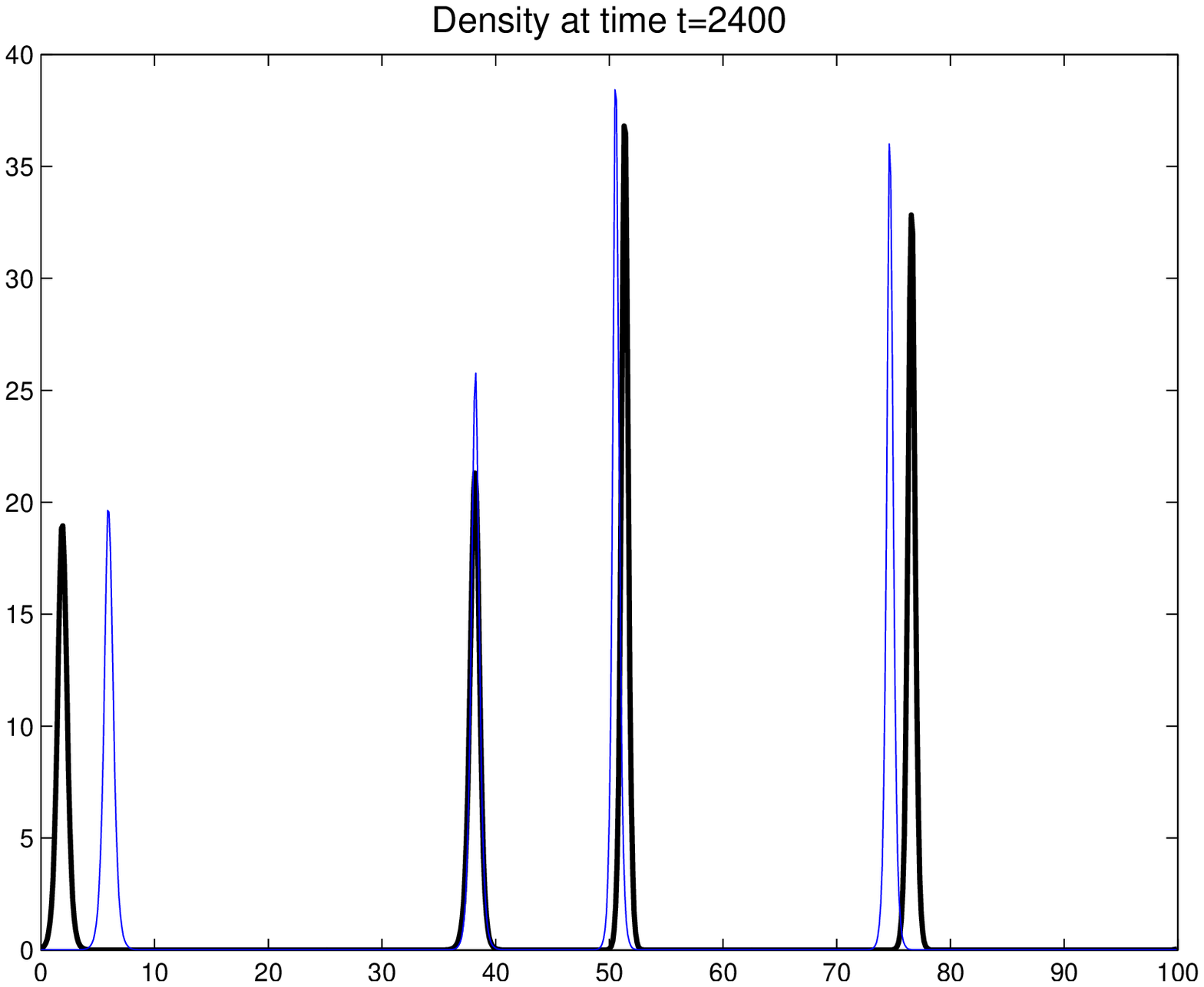}

\includegraphics[width=0.4\textwidth]{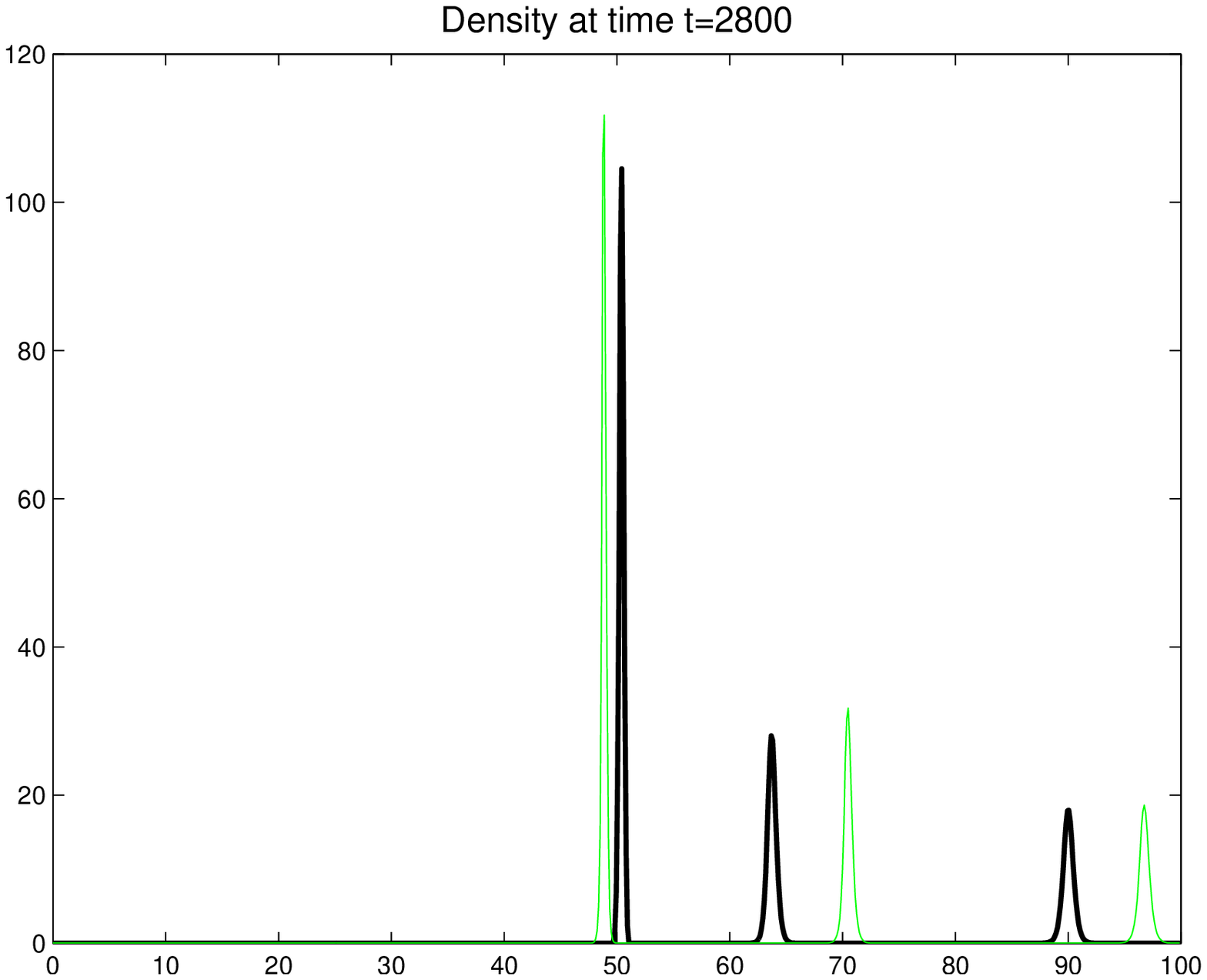}\hfill
\includegraphics[width=0.4\textwidth]{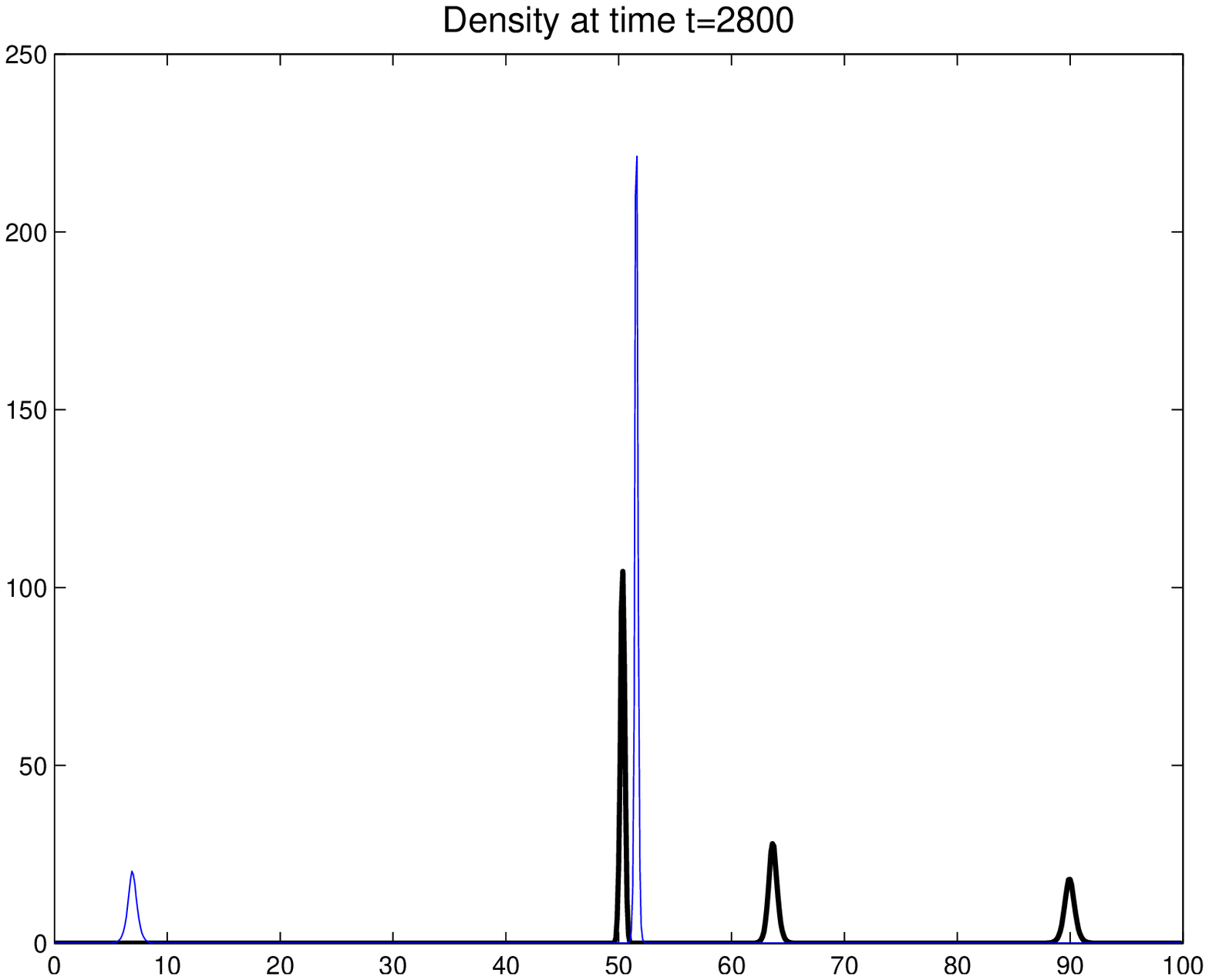}

\includegraphics[width=0.4\textwidth]{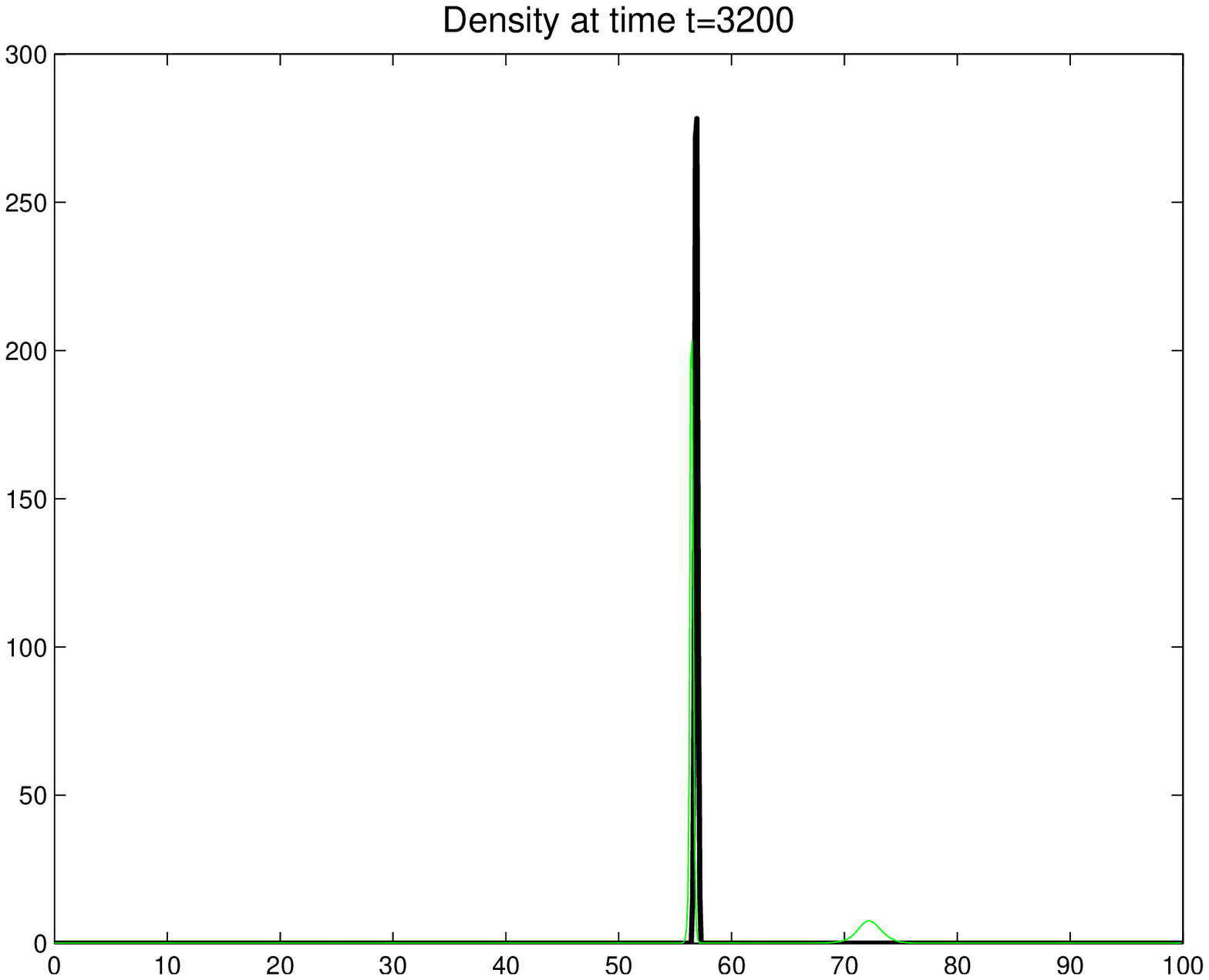}\hfill
\includegraphics[width=0.4\textwidth]{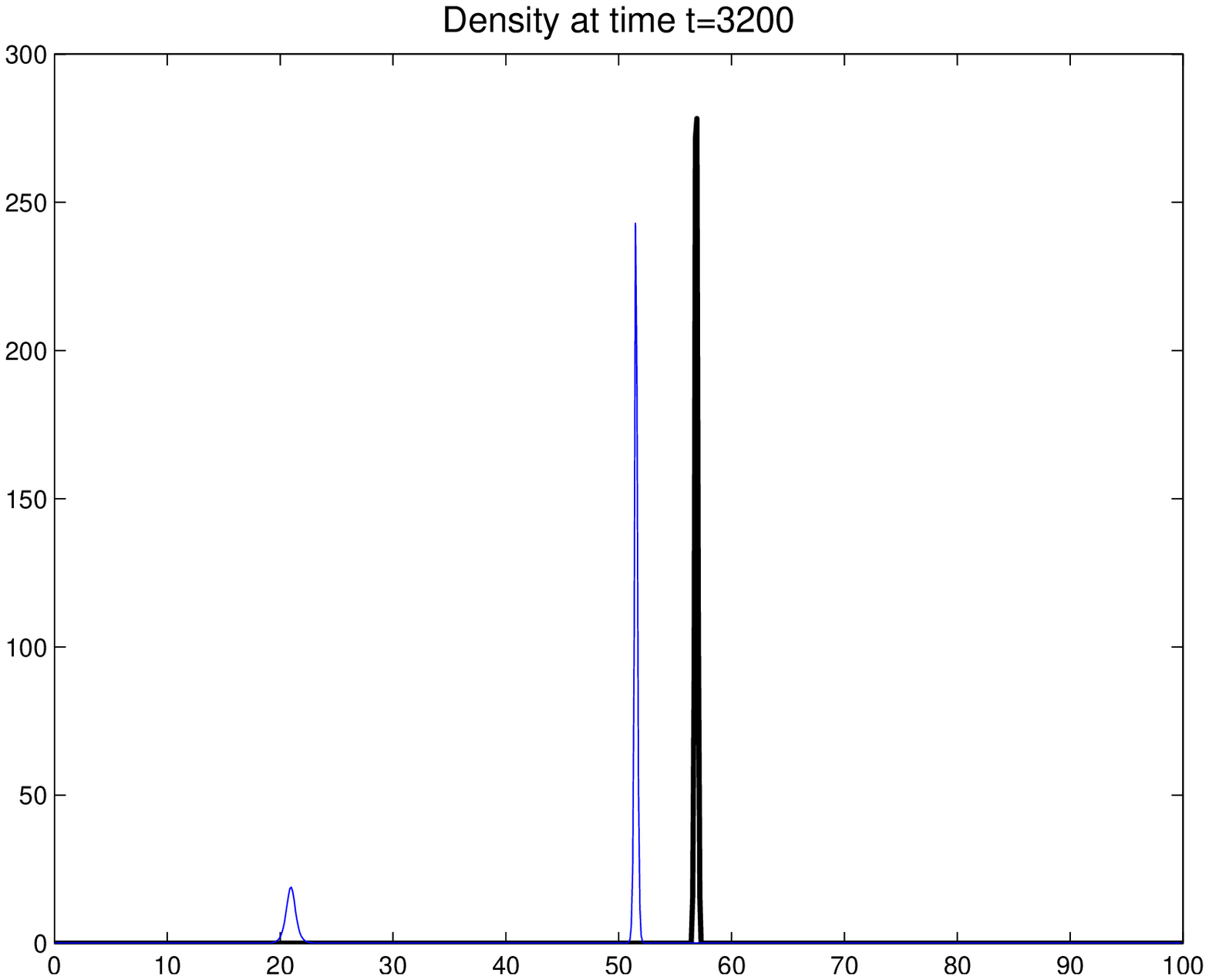}
\caption{\label{fig:numerics-3}
Results of a simulation with $\delta = 0.2$ at later times.
Left plots: Comparison of the quasi-dynamic approximation (green line) with
  the full model (black line). Right plots: Comparison of the
  diffusive scaling (blue line) with the full model (black line).}
\end{figure}

At later times the cluster start to merge. This coarsening behavior
can be observed with all of the three models. Here a better agreement
is observed for the quasi-dynamic approximation, see Figure \ref{fig:numerics-3}.

\section*{Conclusions}
Based on a simplified kinetic model, we have studied the sedimentation
of rod-like particles under the influence of gravity.
Linear stability shows both instability of a well stirred initial
configuration as well as a wave length selection mechanism for a 
non-zero Reynolds number.
We presented two models describing the macroscopic response
of the system. One of these models, the quasi-dynamic approximation,
is obtained from a moment closure system using the assumption that
the evolution equation of the second order moments can be replaced by
an equilibrium relation. The resulting macroscopic system has the form
of a flux-limited Keller-Segel model.  
Another macroscopic model is obtained by taking the diffusive limit of
the kinetic model. In this case we obtain a standard Keller-Segel type
model. 
Numerical computations confirm good agreement of the predicted
solution structure of both macroscopic models compared to the kinetic
model. For very long times the quasi-dynamic approximation shows a 
better agreement with the original kinetic model.

Finally, it is interesting to note the differences of the macroscopic
models depending on whether the orientation of the rod-like particles
is a function on $S^1$ or $S^2$, respectively.
The macroscopic models which are derived from the
simplified kinetic model are less diffusive than those which are
derived from the more general kinetic model.

\begin{appendix}
\section{Collective behavior in scaling limits} \label{app:A}
In this appendix we derive the hyperbolic and diffusive limits
for the system (\ref{eqn:system_nondim}).  The description of collective behavior 
of kinetic models through hyperbolic or parabolic limits is well known in several contexts
in fluid dynamics or biological transport systems ({\it e.g.} \cite{CMPS04,OH02,PS00}).
The novelty here is that the kinetic variable takes values in the sphere, what requires
some special calculations detailed in this appendix.

It is expedient to view the scaling limits from the perspective 
of describing the aggregate behavior of a suspension.  The function
$$
\rho(t, \vec{x}) = \int_{S^{d-1}} f(t, \vec{x},\vec{n}) d \vec{n}
$$
measures the density of rod-like particles.
Linear stability theory predicts an instability for the quiescent solution;
it is then natural to calculate the aggregate response of the system  in long times. 
To this end, we proceed to calculate the hyperbolic and diffusive limits.
It turns out that the limiting behavior in the hyperbolic scaling will be described 
by a Boussinesq type system. For certain flows the hyperbolic  scaling 
produces a trivial behavior, and it is then natural to consider the diffusive scaling.
Such a situation occurs for two-dimensional rectilinear flows of suspensions,
where we will show that the collective behavior in the diffusive limit is
described by the Keller-Segel model.

\subsection{The hyperbolic scaling}

We first rescale the model \eqref{eqn:system_nondim} 
in the {\em hyperbolic scaling},
$$
\vec{x} = \frac{1}{\delta} \hat{\vec{x}}, \ t = \frac{1}{\delta} \hat{t}, \
\vec{u} = \hat{\vec{u}}, \ p = \hat{p}.
$$
The scaled equations (after dropping the hats) are
\begin{equation}
\label{hypscale}
\begin{aligned}
\delta \partial_{t} f + \delta \vec{u} \cdot \nabla_{\vec x} f
&- \delta D(\vec n)  \vec{e}_2 \cdot \nabla_{\vec x} f  
+ \delta \nabla_{\vec n} \cdot \left( P_{\vec n^\bot} \nabla_{\vec x} \vec{u} \vec n f \right)
\\
 &= \Delta_{\vec n} f + \delta^2\gamma \nabla_{\vec x} \cdot D(\vec n) \nabla_{\vec x} f
  \\
  Re \, \delta \,  \left(\partial_{t} \vec{u} + \left( \vec{u} \cdot \nabla_{\vec{x}} \right) \vec{u}\right)
-\delta^2 \Delta_{\vec x} \vec{u} &+ \delta \nabla_{\vec x} p 
- \delta^2 \gamma \nabla_{\vec x} \cdot \sigma  = \delta
\left(\brho - \int_{S^2} f d \vec n \right) \vec e_3
\\
& \qquad \qquad  \delta \nabla_{\vec x} \cdot \vec{u} = 0
\end{aligned}
\end{equation}
where $D(\vec{n}) = I +\vec{n} \otimes \vec{n}$. 

We introduce the ansatz
\begin{equation*}
\begin{split}
f & =  f_0 + \delta f_1 + \ldots\\
\vec{u} & = \vec{u}_0 + \delta \vec{u}_1 + \ldots
\\
p & =  p_0 + \delta p_1 + \ldots 
\end{split}
\end{equation*}
to the system \eqref{hypscale} and obtain equations for the various
orders of the expansion:
\begin{align}
&O(1)  \quad 
&&\Delta_{\vec n} f_0 = 0
\label{ord1}
\\
&O(\delta)  \quad 
&&\partial_{t} f_0 + \vec{u}_0 \cdot \nabla_{\vec x}  f_0 
- D(\vec n)  \vec e_2 \cdot \nabla_{\vec x} f_0 
+ \nabla_{\vec n} \cdot 
\left( P_{\vec n^\bot}  \nabla_{\vec{x}} \vec{u}_0 \vec{n} f_0  \right ) = \Delta_{\vec n} f_1
\label{ord2}
\\
&O(\delta)  \quad 
&& Re \,  \left(\partial_{t} \vec{u}_0 + \left( \vec{u}_0 \cdot \nabla_{\vec{x}} \right) \vec{u}_0 \right)  +  \nabla_{\vec x} p_0
= \Big (\brho -  \int_{S^2} f_0 d \vec n  \Big ) \vec{e}_3
\nonumber
\\
&O(1)  \quad 
&& \nabla_{\vec x} \cdot \vec{u}_0  = 0
\nonumber
\end{align}
It follows from \eqref{ord1} that $f_0$ is independent of $\vec{n}$ and thus
$$
f_0 (t, \vec{x}, \vec{n}) = \frac{1}{4\pi} \int_{S^2} f_0  d\vec{n}
= \frac{1}{4\pi} \rho_0 (t, \vec{x}) 
$$
Then integrating \eqref{ord2} over the sphere, we deduce that 
$\rho_0 = \int_{S^2} f_0 d\vec n$ and $\vec{u}_0$ satisfy
the Boussinesq system
\begin{equation}
\label{Boussinesq}
\begin{aligned}
\partial_t \rho_0 + \nabla_{\vec x}  \cdot \left ( \vec{u}_0 \rho_0 -
 \big ( \frac{1}{4\pi} \int_{S^2} D (\vec{n}) d\vec{n} \big ) 
\vec{e}_2 \rho_0 \right ) &= 0
\\
 Re \,  \left(\partial_{t} \vec{u}_0  + \left( \vec{u}_0 \cdot \nabla_{\vec{x}} \right) \vec{u}_0 \right)  +  \nabla_{\vec x} p_0
&= (\brho - \rho_0 ) \vec e_3
\\
\nabla_{\vec x} \cdot \vec{u}_0  &= 0
\end{aligned}
\end{equation}
The constant $\brho$ can be absorbed into the pressure, by redefining $p_0$ to account for the hydrostatic pressure, 
that is by setting $p_0 = \hat p_0 + \brho \vec{x} \cdot \vec e_3$.

\subsection{The diffusive scaling}\label{app:diff}

Next, we confine to rectilinear flows with a vertical  velocity field obeying the ansatz
\begin{equation}
\label{ansrecti}
\vec{u}(t,x,y) = (0, 0, w(t,x,y))^T \; , \quad f = f (t, x, y)
\end{equation}
and depending only on the horizontal variables. The flow cross section is the domain $D$ and
we assume that the boundary conditions are either periodic or no-slip.
This restriction  to the two-dimensional case   is motivated by experimental observations of long clusters with higher 
particle density. We note that for this ansatz the nonlinear transport terms $u\cdot \nabla f$ and  $(u \cdot \nabla) u$ drop out.

One checks that under the ansatz \eqref{ansrecti} the system \eqref{Boussinesq} reduces to the trivial problem
$$
\del_t \rho_0 = 0,   \quad  Re \,  \del_t w_0 = (\brho - \rho_0) \, , \quad p_0 = - \brho z
$$
which can be easily solved in terms of the initial data.
The objective then becomes to  calculate the next order correction in the diffusive scale.

We return to \eqref{eqn:system_nondim}, note that for the ansatz
\eqref{ansrecti} the 
nonlinear convective terms
$u\cdot \nabla f$ and  $(u \cdot \nabla) u$ drop out, and 
rescale the model in the {\em diffusive scaling}, i.e.\
\begin{equation}
\label{diffsc}
\vec{x} = \frac{1}{\delta} \hat{\vec{x}}, \ t = \frac{1}{\delta^2} \hat{t}, \
\vec{u} = \hat{\vec{u}}, \ p =  \hat{p}.
\end{equation}
The scaled equations (after dropping the hats) have the form
\begin{equation*}
\begin{split}  
\delta^2 \partial_{t} f 
- \delta  D(\vec n)  \vec e_3 \cdot \nabla_{\vec x} f  
+ \delta \nabla_{\vec n} \cdot \left( P_{\vec n^\bot} \nabla_{\vec x}
  \vec{u} \vec n f \right)  
  &= D_r \Delta_{\vec n} f + \delta^2\gamma \nabla_{\vec x} \cdot D(\vec n) \nabla_{\vec x} f
  \\
\delta^2 \, Re \,\partial_t \vec{u}  + \delta \nabla_{\vec x} p 
&= \delta^2
\Delta_{\vec x} \vec{u} + \delta^2 \gamma \nabla_{\vec x} \cdot \sigma 
+ \delta \Big( \brho - \int_{S^2} f d \vec n \Big) \vec e_3
\\
 \nabla_{\vec x} \cdot \vec{u}  & = 0
\end{split}
\end{equation*}
We introduce the ansatz
\begin{equation}\label{eqn:ansatz-u}
\begin{split}
f (t,x,y, \vec{n}) & = \delta f_0 + \delta^2 f_1 + \ldots\\
\vec{u}(t,x,y)  & = \vec{u}_0 + \delta \vec{u}_1 + \ldots
 = \left(\begin{array}{c}
0\\0\\w_0\end{array}\right) + \delta \left(\begin{array}{c}
0\\0\\w_1\end{array}\right) + \ldots\\
p & = \delta p_0 + \delta^2 p_1 + \ldots  \\
\brho & = \delta \brho_0 + \delta^2 \brho_1 + \ldots
\end{split}
\end{equation}
to the above system and collect the terms of the same order,
arriving at
\begin{align}
&O(\delta)  \quad 
&&\Delta_{\vec n} f_0 = 0
\label{as1}
\\
&O(\delta^2)  \qquad 
&&  - D(\vec n)  \vec e_3
 \cdot \nabla_{\vec x} f_0 
+ \nabla_{\vec n} \cdot 
\left( P_{\vec n^\bot}  \nabla_{\vec{x}} \vec{u}_0 \vec{n} f_0  \right
) = D_r \Delta_{\vec n} f_1
\label{as2}
\\
&O(\delta^3)  \quad
&&\partial_{t} f_0 - D(\vec n)  \vec e_3 \cdot \nabla_{\vec x} f_1
+ \nabla_{\vec n} \cdot 
\left( P_{\vec n^\bot} \big ( \nabla_{\vec{x}} \vec{u}_1 \vec{n} f_0 
+ \nabla_{\vec{x}} \vec{u}_0 \vec{n} f_1 \big ) \right ) 
\nonumber
\\
&
&&\qquad \qquad = D_r \Delta_{\vec n} f_2
+ \gamma \nabla_{\vec{x}} \cdot D(\vec{n}) \nabla_{\vec{x}} f_0
\label{as3}
\end{align}
The same procedure applied to the Stokes system yields
\begin{align}
&O(\delta)  \qquad   && \nabla_{\vec x} \cdot \vec{u}_0  = 0 \nonumber \\
&O(\delta^2)  \qquad 
&& \partial_t \vec{u}_0 +  \nabla_{\vec x} p_0
= \Delta_{\vec x} \vec{u}_0  + \left ( \brho_0  - \int_{S^2} f_0 d \vec n  \right ) \vec{e}_3
\label{ans2} 
\end{align}

We want to derive an evolution equation for $\rho_0$ and $w_0$. 
In order to do this, we first summarise a few tools.
Recall that 
$\vec{n} \in S^2$ has the form
$$\vec{n} = \left( \begin{array}{c}
\sin \theta \cos \phi \\ \sin \theta \sin \phi \\ \cos
\theta \end{array}\right), \quad 0 \le \theta < \pi, \, 0\le \phi < 2\pi.
$$
Furthermore, recall that the  components of the tensor 
$3\vec{n}\otimes \vec{n}-{\rm id}$
are the surface spherical harmonics of order 2. This means, they are
harmonic polynomials on $\mathbb{R}^3$ of order 2, restricted
to $S^2$. The surface spherical harmonics
are eigenfunctions of the Laplacian on $S^2$ with corresponding
eigenvalue $-\ell(\ell+1)$, where $\ell$ is the order
\cite[App. E]{book:BCAH87}. Hence
\begin{equation}\label{surfhar}
\triangle_{\vec{n}} (3\,n_i\,n_j-\delta_{ij})\;=\;-6\,(3\,n_i\,n_j-\delta_{ij}).
\end{equation}
Finally, note that  for any $3\times 3$ matrix $\kappa$, the equation 
\begin{equation}\label{A.5}
\nabla_{ \vec{n}} \cdot \left( P_{\vec{n}^\bot} \kappa \vec{n} \right) 
= \mbox{tr } \kappa - 3 \vec{n} \cdot \kappa \vec{n} \, ,
\end{equation} 
holds, where tr stands for the trace operator. Also, using symmetries of $S^2$, we obtain the formula
\begin{equation}
\label{form1}
\frac{1}{4 \pi} \int_{S^2} \vec n \otimes \vec n d\vec n = \tfrac{1}{3} I .
\end{equation}

Now we are ready to derive an evolution equation for $\rho_0$.
Equation \eqref{as1} implies that $f_0$ is independent on $\vec{n}$, that is
$$
f_0 = \frac{1}{4\pi} \int_{S^2} f_0 d\vec{n} = \frac{1}{4\pi} \rho_0 (t,x,y)
$$
Next,  integration of \eqref{as3} over the sphere and use of \eqref{form1} and the fact that $e_3 \cdot \nabla_x f_1 = 0$ for our ansatz  gives that $\rho_0$ satisfies
\begin{equation} \label{eqn:KS1}
\partial_t \rho_0 = \nabla_{\vec x} \cdot  \underbrace{
 \int_{S^2} \left( D(\vec n) - \frac{1}{4 \pi} \int_{S^2} D(\vec n) d \vec n \right) \vec e_3  f_1 \, d\vec n}_{=: I_1} 
+ \gamma \nabla_{\vec x} \cdot
\underbrace{ \frac{1}{4 \pi} \int_{S^2} D(\vec n) d \vec n
\,}_{=:I_2}
\nabla_{\vec x} \rho_0 
\end{equation}
where the terms $I_1$ and $I_2$ are computed in terms of $f_1$ solving \eqref{as2} for
$f_0 = \frac{1}{4\pi} \rho_0$.

It remains to compute the terms $I_1$ and $I_2$.
Observe now that we have the identities: 
\begin{align}
\int_{S^2} ( \vec{n}\otimes \vec{n} - \frac{1}{3} I ) d\vec{n} &\stackrel{\eqref{surfhar}}{=}
-\frac{1}{6} \int_{S^2} \triangle_{\vec{n}} ( \vec{n}\otimes \vec{n} - \frac{1}{3} I ) d\vec{n}
= 0
\nonumber
\\
I_2 := \frac{1}{4\pi} \int_{S^2} D(\vec{n}) d\vec{n} &= \frac{1}{4\pi} \int_{S^2} 
( \vec{n}\otimes \vec{n} + I  )  d\vec{n} = \frac{4}{3} I
\\
D(\vec{n}) - \frac{1}{4\pi} \int_{S^2} D(\vec{n}) d\vec{n} &= 
\vec{n}\otimes \vec{n} - \frac{1}{3} I
\nonumber
\end{align}
These, in conjunction with \eqref{as1}, \eqref{as2} and \eqref{A.5}, 
imply that $f_1$ satisfies
\begin{align}
\Delta_{\vec n} f_1 &= 
- \big ( D(\vec n) - \frac{1}{4\pi} \int_{S^2} D(\vec n) d\vec{n}  \big ) \vec e_3
 \cdot \frac{1}{4\pi} \nabla_{\vec x} \rho_0 
+ \frac{1}{4\pi} \nabla_{\vec n} \cdot 
\left( P_{\vec n^\bot}  \nabla_{\vec{x}} \vec{u}_0 \vec{n} \right )\rho_0
\nonumber
\\
&= - \big ( \vec{n}\otimes \vec{n} - \frac{1}{3} I   \big ) \vec e_3
 \cdot \frac{1}{4\pi} \nabla_{\vec x} \rho_0 
- \frac{3}{4\pi} \rho_0 
( \vec{n} \cdot \nabla_{\vec{x}} \vec{u}_0 \vec{n} )
\label{as4}
\end{align}

Next, we compute $I_1$
\begin{equation*}
\begin{split}
I_1 &:=  \int_{S^2} \left( D(\vec n) - \frac{1}{4 \pi} \int_{S^2}
  D(\vec n) d \vec n \right) 
\vec e_3  f_1 \, d\vec n
\\
&= \int_{S^2} \big ( \vec{n}\otimes \vec{n} - \frac{1}{3} I   \big ) \vec e_3 f_1 d\vec{n}
\\
&\stackrel{\eqref{surfhar}}{=} -\frac{1}{6} 
 \int_{S^2} \triangle_{\vec{n}} \big ( \vec{n}\otimes \vec{n} - \frac{1}{3} I   
 \big ) \vec e_3 f_1 d\vec{n}
\\
&\stackrel{\eqref{as4}}{=}
\frac{1}{24\pi}  \int_{S^2} \big ( \vec{n}\otimes \vec{n} - \frac{1}{3} I   \big ) \vec e_3 
\Big [ \big ( \vec{n}\otimes \vec{n} - \frac{1}{3} I   \big ) \vec e_3 \cdot \nabla_{\vec{x}}\rho_0
+ 3 \rho_0 ( \vec{n} \cdot \nabla_{\vec{x}} \vec{u}_0 \vec{n} )  \Big ] d\vec{n}
\\
&= \frac{1}{24\pi}  \int_{S^2} 
\left( \begin{array}{c}
n_1 n_3 \\
n_2 n_3 \\
n_3^2 - \frac{1}{3}  \end{array} 
\right )
\Big [ (3\rho_0  w_{0x} + \rho_{0x}) n_1 n_3 + (3\rho_0  w_{0y} + \rho_{0y}) n_2 n_3 
\Big ] d\vec{n}
\end{split}
\end{equation*}

Observe that, due to symmetry considerations, the integrals
$$
\begin{aligned}
\int_{S^2} n_1 n_3^3 - \frac{1}{3} n_1 n_3 d\vec{n} &= 0
\\
\int_{S^2} n_2 n_3^3 - \frac{1}{3} n_2 n_3 d\vec{n} &= 0
\\
\int_{S^2} n_1 n_2 n_3^2 d\vec{n} = &=0,
\end{aligned}
$$
while the remaining integrals are computed via spherical coordinates
$$
\begin{aligned}
\int_{S^2} n_1^2 n_3^2 d\vec{n} &= \int_0^\pi \sin^3 \theta \cos^2 \theta \,  d\theta 
\int_0^{2\pi} \cos^2 \varphi d\varphi = \frac{4\pi}{15}
\\
\int_{S^2} n_2^2 n_3^2 d\vec{n} &= 
\int_0^\pi \sin^3 \theta \cos^2 \theta \,  d\theta 
\int_0^{2\pi} \sin^2 \varphi d\varphi = \frac{4\pi}{15}. 
\end{aligned}
$$
We conclude that
$$
I_1 = \frac{1}{90} \left ( 3 \rho_0  w_{0x} + \rho_{0x} \, , \;  3
  \rho_0  w_{0y} + \rho_{0y} \, , 0\;\right )^T
$$
and that $\rho_0$ satisfies the equation
\begin{align*}
\partial_t \rho_0 &= \frac{1}{30}  \left(
\partial_x \left(\frac{1}{3} \rho_{0x} + \rho_0 w_{0x} \right) + \partial_y \left( \frac{1}{3} \rho_{0y} + \rho_0 w_{0y} \right) \right) + \gamma
\frac{4}{3} \Delta_{(x,y)} \rho_0 
\\
&= \frac{1}{30} \nabla_{(x,y)} \cdot \big ( \rho_0 \nabla_{(x,y)} w_0 \big )
+ \frac{1}{3} \left ( 4\gamma + \frac{1}{30} \right ) \triangle_{(x,y)} \rho_0
\end{align*}

Finally, we want to derive the evolution equation for $w_0$. 
From  (\ref{ans2}) we obtain
$$
\del_x p_0 = \del_y p_0 = 0
$$
and
\begin{equation}\label{eqn:p1}
 Re \partial_t w_0  -  \Delta_{(x,y)}  w_0  + \del_z p_0 = (\brho_0 - \rho_0 )
\end{equation}
The ansatz (\ref{eqn:ansatz-u}) implies that the right hand side of
(\ref{eqn:p1}) depends only on $(x,y)$. Thus we deduce that the
pressure has the form
$p = \kappa(t) z$, where $\kappa$ is arbitrary and reflects the effect
of an imposed pressure gradient. If there is no imposed pressure gradient then the pressure is hydrostatic
and $\brho_0 =  \tfrac{1}{|D|} \int_D \rho (t,x,y) dx dy$ which is conserved.
In the following we restrict our
considerations to the case $\gamma = 0$. The functions $(\rho_0, w_0)$
are selected by solving the coupled system
\begin{equation}
\begin{split}
\partial_t \rho_0 & = \frac{1}{30} \nabla_{(x,y)} \cdot \left( \rho
  \nabla_{(x,y)} w_0 \right) + \frac{1}{90} \Delta_{(x,y)} \rho_0 \\
Re \, \partial_t w_0 & = \Delta_{(x,y)} w_0 + (\brho -\rho_0),
\end{split}
\end{equation}
where $\brho$ is (as above) the average density.

\end{appendix}



\begin{thebibliography}{1}

\bibitem{book:BCAH87}
R.B. Bird, Ch.F. Curtiss, R.C. Armstrong, and O.~Hassager.
\newblock {\em Dynamics of Polymeric Liquids, Vol. 2, Kinetic Theory}.
\newblock Wiley Interscience, 1987.

\bibitem{CGL08}
J.-A. Carillo, Th. Goudon and P. Lafitte.
\newblock Simulation of fluid and particles flows: 
Asymptotic preserving schemes for bubbling and flowing regimes.
\newblock {\em J. Comput. Physics}, 227 (2008) 7929-7951.

\bibitem{CMPS04}
F.A.C.C. Chalub, P.A. Markowich, B.~Perthame, and C.~Schmeiser.
\newblock Kinetic models for chemotaxis and their drift-diffusion limits.
\newblock {\em Monatsh.\ Math.}, 142 (2004), 123--141.

\bibitem{book:DE86}
M.~Doi and S.F. Edwards.
\newblock {\em The Theory of Polymer Dynamics}.
\newblock Oxford University Press, 1986.

\bibitem{GH11}
\'E.~Guazzelli and J.~Hinch.
\newblock Fluctuations and instability in sedimentation.
\newblock {\em Annu.\ Rev.\ Fluid Mech.}, 42 (2011), 97-116.

\bibitem{HT2015}
Ch. Helzel and A.E. Tzavaras.
\newblock  A kinetic model for the sedimentation of rod-like particles.
\newblock 2015 (submitted).

\bibitem{HG96}
B.~Herzhaft and \'E.~Guazzelli.
\newblock Experimental study of the sedimentation of a dilute fiber
suspension.
\newblock {\em Physical Review Letters}, 77 (1996), 290-293.

\bibitem{HG99}
B.~Herzhaft and \'E.~Guazzelli.
\newblock Experimental study of the sedimentation of dilute and
semi-dilute suspensions of fibers.
\newblock {\em J.\ Fluid Mech.}, 384 (1999), 133-158.

\bibitem{KS89}
D.L.~Koch and E.S.G.~Shaqfeh.
\newblock The instability of a dispersion of sedimenting spheroids.
\newblock {\em J.\ Fluid Mech.}, 209 (1989), 521-542.

\bibitem{MBG07}
B.~Metzger, J.E.~Butlerz and \'E.~Guazzelli.
\newblock Experimental investigation of the instability of a sedimenting suspension
of fibers.
\newblock {\em J.\ Fluid Mech.}, 575 (2007), 307-332.

\bibitem{OH02}
H.G. Othmer and T. Hillen.
\newblock The diffusion limit of transport equations II: Chemotaxis equations.
\newblock {\em SIAM J. Appl. Math.} {\bf 62} (2002), 1222-1250.

\bibitem{PS00} F. Poupaud and J. Soler.
\newblock Parabolic limit and stability of the Vlasov-Poisson-Fokker-Planck system.
\newblock {\em  Math. Mod. Meth. Appl. Sci.} {\bf 10}  (2000), 1027Ð1045.

\end{thebibliography}

\end{document}